\documentclass[9pt, oneside]{article}
\usepackage[T1]{fontenc}
\usepackage[utf8]{inputenc}
\usepackage{setspace}
\usepackage[a4paper, width=170mm,top=25mm,bottom=25mm]{geometry}
\usepackage{graphicx}
\graphicspath{{images/}}
\usepackage{wrapfig}
\usepackage{booktabs}

\usepackage{multicol}

\usepackage{hyperref}
\usepackage[title]{appendix}
\usepackage[super,sort&compress,comma]{natbib}

\hypersetup{colorlinks, linkcolor = black , urlcolor = blue, citecolor = blue}

\usepackage[table,xcdraw]{xcolor}

\usepackage{amsmath,amsfonts}
\usepackage{mathrsfs}
\usepackage[version=3]{mhchem}
\usepackage{mathptmx, bm}
\usepackage{lastpage}
\usepackage{float}
\usepackage{fancyhdr}
\usepackage{fnpos}
\usepackage[english]{babel}

\emergencystretch=\maxdimen
\usepackage{caption}

\usepackage{etoolbox}
\makeatletter
\patchcmd{\@makechapterhead}{50\p@}{0pt}{}{}
\patchcmd{\@makeschapterhead}{50\p@}{0pt}{}{}
\makeatother
\usepackage{titlesec}

\AtBeginDocument{%
  \setlength{\bibsep}{0.0pt}
  \titlespacing{\section}{0pt}{2pt plus 1pt minus 1pt}{2pt plus 1pt minus 1pt}
  \titlespacing{\subsection}{0pt}{2pt plus 1pt minus 1pt}{2pt plus 1pt minus 1pt}
  }

\allowdisplaybreaks
\usepackage{pdfpages}

\usepackage{pifont}% http://ctan.org/pkg/pifont
\newcommand{\cmark}{\ding{51}}%
\newcommand{\xmark}{\ding{55}}%
\usepackage{authblk}

\begin{document}
\captionsetup{font={small,singlespacing}}
\newcommand \e[1]{{\textbf{#1}}}
\newcommand \bo[1]{{\bf{{#1}}}}
\newcommand \bra[1] {\left\langle {#1} \right\vert}
\newcommand \ket[1] {\left\vert {#1} \right\rangle }
\newcommand \braket[2] {\left\langle {#1} \vert{#2} \right\rangle}

\title{\large{Which Algorithm Best Propagates the Meyer-Miller-Stock-Thoss Mapping Hamiltonian for Non-Adiabatic Dynamics?}}
\author[1]{\normalsize{Lauren E. Cook}}
\author[2]{Johan E. Runeson \footnote{ Current Address: Department of Chemistry, University of Oxford, Physical and Theoretical Chemistry Laboratory, South Parks Road, Oxford OX1 3QZ, United Kingdom}}
\author[2]{Jeremy O. Richardson}
\author[1]{Timothy J. H. Hele \footnote{Corresponding Author: \href{mailto:t.hele@ucl.ac.uk}{t.hele@ucl.ac.uk}}}

\affil[1]{Department of Chemistry, University College London, Christopher Ingold Building, London WC1H 0AJ, United Kingdom}
\affil[2]{Department of Chemistry and Applied Biosciences, ETH Zürich, 8093 Zürich, Switzerland}

\date{\normalsize{\today}}
\maketitle

A common strategy to simulate mixed quantum-classical dynamics is by propagating classical trajectories with mapping variables, often using the Meyer--Miller--Stock--Thoss (MMST) Hamiltonian or the related spin-mapping approach. 
When mapping the quantum subsystem, the coupled dynamics reduce to a set of equations of motion to integrate. 
Several numerical algorithms have been proposed, but a thorough performance comparison appears to be lacking. 
Here, we compare three time-propagation algorithms for the MMST Hamiltonian: the Momentum Integral (MInt) (J. Chem. Phys., 2018, 148, 102326), the Split-Liouvillian (SL) (Chem. Phys., 2017, 482, 124–134), and the algorithm in J. Chem. Phys., 2012, 136, 084101 that we refer to as the Degenerate Eigenvalue (DE) algorithm due to the approximation required during derivation.
We analyse the accuracy of individual trajectories, correlation functions, energy conservation, symplecticity, Liouville's theorem and the computational cost. We find that the MInt algorithm is the only rigorously symplectic algorithm. 
However, comparable accuracy at a lower computational cost can be obtained with the SL algorithm. 
The approximation implicitly made within the DE algorithm conserves energy poorly, even for small timesteps, and thus leads to slightly different results.
These results should guide future mapping-variable simulations.  
\begin{multicols}{2}

\section{Introduction}
Theoretical methods for simulating nonadiabatic dynamics are crucial for understanding charge and energy transfer in materials such as organic light-emitting diodes (LEDs), solar cells and photosynthetic systems.\cite{Church2018, Kelly2012, Richardson2017,Reece2009, Domcke2012,Hammes-Schiffer2010,Cheng2009} Experimental spectroscopic techniques have successfully been developed to probe nonadiabatic processes within photovoltaic materials.\cite{DeSio2018, Nguyen2019, Gaynor2018} To analyse the complex dynamics observed, accurate and efficient numerical models are of great importance.\cite{Richardson2017, Nguyen2019} 

Since the 1930s, when Landau and Zener first investigated nonadiabatic dynamics,\cite{Landau1932,Zener1932} accurate calculation of nonadiabatic dynamics has been a challenge due to the large computational expense associated with full-quantum solutions.\cite{Hele2016,Topaler1996, Beck2000} This led to the development of approximate classical-like dynamical frameworks, motivated by the classical linear scaling with degrees of freedom (DoF) compared to the exponential scaling of quantum methods.\cite{Stock2005} Unlike most transformations, discrete quantum DoF do not have an obvious classical counterpart.\cite{Stock2005} 
Consequently, approximate methods have been developed to recast the quantum system to look classical whilst retaining some quantum properties, often requiring a compromise between accuracy and cost for large systems.

Many methods exist to incorporate discrete quantum DoF into classical frameworks including Ehrenfest dynamics,\cite{Shalashilin2011} approximations to the quantum-classical Liouville equation (QCLE),\cite{Kapral2006, Donoso1998, Kapral2016} the symmetrical quasi-classical windowing method,\cite{Cotton2016} surface hopping,\cite{Tully1971, Tully1990} %Zimmermann2014} 
and mapping methods such as methods inspired by the Meyer--Miller--Stock--Thoss (MMST) mapping,\cite{Huo2011, Meyer1979, Stock1997, Wang1999, Antipov2015} and spin-mapping,\cite{Runeson2019, Runeson2020} including the mapping approach to surface hopping (MASH).\cite{MASH}  The simplest nonadiabatic method, proposed by Mott (1931), evaluates the quantum electronic dynamics along the classical path of the nuclei, known as the ‘classical path’ approach.\cite{Stock2005, Mott1931, Kapral2006, Donoso1998} 
However, this does not include the `back reaction' on the nuclei.\cite{Stock2005}
Surface hopping models, originally developed by Tully \textit{et al.}~(1971), propagate along one adiabatic surface before ‘hopping’ to another.\cite{Stock2005, Tully1990, Tully1971, Prezhdo1997, Zimmermann2014} ‘Hopping’ is possible at any point along a trajectory, not just where surfaces cross, destroying the state coherence.\cite{Stock1997, Tully1990}

In this article, we focus on mapping approaches, utilised in various semiclassical methods, where the discrete quantum DoF are mapped onto continuous classical DoF propagated by classical mechanics. %\cite{Hele2016, Stock1997}
The Meyer--Miller mapping developed in 1979,\cite{Meyer1979} and later put on a rigorous footing by Stock and Thoss in 1997,\cite{Stock1997} constructs a set of classical variables for the discrete electronic DoF and propagates them with the nuclear DoF using an effective Hamiltonian, the MMST Hamiltonian.\cite{Stock1997, Meyer1979, Meyer1979b} The electronic dynamics are consistent with the time-dependent Schrödinger equation, and the force exerted on the nuclei is given by the instantaneous values of the electronic variables.\cite{Meyer1979, Meyer1979b} The mapping introduces electronic position and momenta, sometimes expressed as action--angle variables, to describe the nuclear motion on coupled potential energy surfaces.\cite{Stock2005, Stock1997} This approach maps the time-dependent Schrödinger equation for an \textit{N}-level system to a classical analog of \textit{N} coupled harmonic oscillators following Hamilton’s equations of motion.\cite{Stock1997, Stock2005} 

The recently derived spin-mapping approach employs a different mapping formalism but gives a Hamiltonian almost identical to that of the MMST methods.\cite{Runeson2019, Runeson2020, Runeson2021} The MMST Hamiltonian algorithms compared in this work are thus also applicable to spin-mapping methods.
Compared to MMST-based methods, spin-mapping uses a different value of the so-called zero-point energy (ZPE) parameter, introduced by Stock and Müller as a fitting parameter to mitigate ZPE-leakage.\cite{Muller1999, Stock1999} Spin-mapping leads to a ZPE as a function of the number of states, with values close to what was previously found optimal when tuned as a free parameter. A partially linearized spin-mapping method has been found to improve accuracy compared to the fully linearized mapping,\cite{Mannouch2020, Mannouch2020b} in particular for spectroscopy.\cite{Mannouch2022}

One application of the MMST Hamiltonian is to calculate dynamical properties at thermal equilibrium by approximating equilibrium time-correlation functions
\begin{align}
    \label{qtcf}
    C_{AB}(t) &= \frac{1}{Z} \mathrm{Tr} \left[ e^{-\beta \hat{H}} \hat{A}(0) \hat{B}(t) \right] \text{,}
\end{align}
where the partition function $Z$ is given as
\begin{align} \label{pf}
    Z &= \mathrm{Tr} \left[ e^{-\beta \hat{H}} \right] \text{,}
\end{align}
and the quantum Boltzmann operator at inverse temperature $\beta = 1/k_{B}T$ is $e^{-\beta \hat{H}}$. Semiclassical methods can be used to calculate quantum time-correlation functions, often by utilising an ‘Initial-Value Representation’ (IVR), resulting in a phase-space integral.\cite{Miller1970,Church2018, Stock1997} 
The semiclassical phase-factor makes correlation function convergence challenging.\cite{Church2018,Bonella2001, Wang1999} Various versions have been developed to attempt to overcome this, including linearised semiclassical (LSC)-IVR,\cite{Church2018, Wang1999, Wang1998, Sun1998a, Sun1998b, Shi2004} and Mixed Quantum-Classical (MQC)-IVR.\cite{Shi2004,Antipov2015, Liu2015,Church2017, Filinov1986, Thoss2001} 
However, semiclassical methods often fail at describing nuclear quantum effects.\cite{Ananth2007}

An alternative method is to use a ring of multiple classical system replicas (beads) attached by harmonic springs, known as path integral molecular dynamics (PIMD).\cite{Richardson2013a, Ananth2013, Feynman1948, Deymier2016} 
While the path-integral representation leads to an exact method for static equilibrium properties, it can also yield short-time approximations to quantum real-time dynamics through methods such as; ring-polymer molecular dynamics (RPMD),\cite{Habershon2013, Craig2004, Hele2015, Richardson2013a} centroid molecular dynamics (CMD),\cite{Cao1993, Cao1994, Cao1994a, Cao1994b} and thermostatted (T)-RPMD.\cite{Rossi2014, Hele2016a, Hele2015c} 
Specifically, RPMD provides, for a single (adiabatic) potential, an approximation to Kubo-transformed time correlation functions while preserving the Boltzmann distribution and is exact in the short-time and classical limits.\cite{Hele2013, Hele2015, Craig2004}  
The Kubo-transformed correlation functions allow short-time quantum effects to be included, although the long-time quantum coherence effects are neglected. RPMD can be derived from exact quantum dynamics through a series of approximations via Matsubara dynamics,\cite{Hele2015, Hele2015b} and RPMD transition-state theory is equivalent to true quantum transition-state theory.\cite{Hele2014, Hele2013, Hele2013b, Althorpe2013} 
Many extensions have been suggested for multiple states, including mean-field RPMD,\cite{Hele2011} nonadiabatic RPMD (NRPMD) and mapping-variable RPMD (MVRPMD).\cite{Richardson2013a, Ananth2013, Richardson2017, Duke2015, Duke2017, Chowdhury2017, Chowdhury2021, Shakib2017}  
However, none of these methods alone fulfils the three important limits of; replicating Rabi oscillations in the uncoupled case, preserving the quantum Boltzmann distribution, and reducing to classical dynamics in the adiabatic limit.\cite{Amati2023, Hele2016, Runeson2019} While a recently developed ellipsoid spin-mapping fulfils all these limits, its mean-field dynamics was found to be often less accurate for short times than the original spin-mapping.\cite{Amati2023} Further work is still required to find an accurate trajectory-based approach to replicate the true quantum dynamics.

Although mapping Hamiltonians have become increasingly popular for simulating non-adiabatic dynamics, numerical integration of the equations of motion is not straightforward due to coupling between the electronic momenta and nuclear positions. Over the years various algorithms have been suggested to solve this problem,\cite{Church2018, Richardson2017,Kelly2012} but as far as we are aware there has been no published computational comparison of MMST algorithms to determine their properties and accuracy. To address this, in this paper we compare the symplectic MInt algorithm by Church \textit{et al.}~(2018), the Split-Liouvillian (SL) algorithm by Richardson \textit{et al.}~(2017), and the algorithm outlined by Kelly \textit{et al.}~(2012), referred to here as the Degenerate Eigenvalue (DE) algorithm.\cite{Church2018, Richardson2017, Kelly2012} Other algorithms exist for this problem, such as Runge--Kutta or the Adams--Bashforth Predictor--Corrector algorithm which are known to be non-symplectic.
We note that one cannot directly use velocity-Verlet due to coupling between the nuclear positions and the mapping momenta.
To avoid integrating stiff equations of motion, Wang \textit{et al.}~suggested an intelligent canonical transformation.\cite{Wang1999}
However, even with this transformation, these approaches are not ideal for propagating mapping variables and will still require short time steps.
We will not consider these algorithms further and instead focus on algorithms which attempt to propagate the mapping variables exactly for an arbitrary timestep.

We seek to compare the algorithms for the simplest possible system for which they can all be compared on an equal footing, and for which there already exists results in the literature for qualitative comparison. To this end, we compute position and state autocorrelation functions for a two-state linear vibronic potential with the MMST Hamiltonian (corresponding to the single-bead limit of NRPMD) using these three algorithms.\cite{Richardson2013a} As many of the methods discussed share similar Hamiltonian forms to the MMST, including spin-mapping, Ehrenfest and some surface hopping models, the results are widely applicable and should inform future computational studies using mapping variable methods.

The article is structured as follows. In section \ref{backgroundtheory} we provide theoretical background for the three algorithms investigated. In section \ref{results} we investigate the symplecticity, satisfaction of Liouville's theorem, accuracy of trajectories and correlation functions, computational cost and energy conservation. We conclude in section \ref{conclusions}.

\section{Background Theory} \label{backgroundtheory}
Here, we present the algebraic forms of the three algorithms on an equal footing, such that we can compare properties including the accuracy, symplecticity and if they satisfy Liouville's theorem.

Note that a brief theoretical analysis of the MInt and SL algorithms, determining the symplecticity in particular, was previously carried out in Ref.~[\!\!\citenum{Church2018}]. We extend the work in Ref.~[\!\!\citenum{Church2018}] to a more thorough computational investigation where we compare the algorithmic performance utilising a model for which there exists results in the literature, and include the DE algorithm. As far as we are aware, the symplecticity has not been rigorously determined for the DE algorithm.

\subsection{Symplectic Integrators}
The Hamiltonian for an \textit{N}-level electronic system in the diabatic representation is 
\begin{equation} \label{ODH}
	H= \frac{1}{2}\bo{p}^T \boldsymbol{\mu}^{-1}\bo{p} + \sum_{n,m=1}^{N} \ket{n} \bo{V}_{nm}(\bo{x}) \bra{m} \text{,}
\end{equation}
where the nuclear position and momenta are \e{x} and \e{p} respectively, \e{V}(\e{x}) is an $N \times N$ diabatic electronic potential energy matrix in the basis of the electronic states, $\ket{n}$, and $\boldsymbol{\mu}$ is a diagonal matrix of nuclear masses. 
The classical MMST mapping Hamiltonian in the diabatic representation is\cite{Meyer1979, Stock1997}
\begin{equation} \label{MMST}
	H= \frac{1}{2} \left\{ \bo{p}^T \boldsymbol{\mu}^{-1} \bo{p}+ \bo{P}^T \bo{V}(\bo{x})\bo{P} + \bo{X}^T \bo{V}(\bo{x})\bo{X} - \mathrm{Tr}[\bo{V}(\bo{x})] \right\} \text{,}
\end{equation}
where \e{X} and \e{P} are the electronic position and momenta respectively.
Classical evolution under this Hamiltonian refers to the equations of motion\cite{Stock1997, Stock2005}
\begin{equation} \label{HEOM} 
    \dot{\bo{X}}(t)	= \frac{\partial H}{\partial \bo{P}} \ \text{,}\   \dot{\bo{P}}(t)	= -\frac{\partial H}{\partial \bo{X}}
    \ \text{,}\  
    \dot{\bo{x}}	= \frac{\partial H}{\partial \bo{p}} \ \text{,}\    \dot{\bo{p}}	= -\frac{\partial H}{\partial \bo{x}} \text{,}
\end{equation}
that constitute a time-dependent Hamiltonian system 
\begin{gather}
    \label{Hsys} 
    \frac{d}{dt}\bo{z} = \bo{J}\nabla_{\bo{z}} H(\bo{z}, t) \text{,} 
\end{gather}
where $\bo{z} = [\bo{x}^\textrm{T}, \bo{X}^\textrm{T}, \bo{p}^\textrm{T}, \bo{P}^\textrm{T}]^\textrm{T}$, the $\nabla_{\bo{z}}$ operator contains the partial derivatives, \e{J} is the structure matrix
\begin{gather}
\label{strucmatrix} 
    \bo{J} = \left[ \begin{matrix}
     \mathbb{O} & \mathbb{I} \\
     -\mathbb{I} & \mathbb{O} \end{matrix} \right] \text{,}
\end{gather}
and $\mathbb{I}/\mathbb{O}$ are the identity and zero matrices respectively.\cite{Church2018, Leimkuhler2005, Meyer1979}
A Hamiltonian integrator is said to be \emph{symplectic} if it fulfils the condition\cite{Church2018, Leimkuhler2005, Hardy1995}
\begin{equation} \label{symplecticity}
\bo{M}^\textrm{T}\bo{J}^{-1}\bo{M} = \bo{J}^{-1} \text{,}
\end{equation}
where $\bo{M}$ is the monodromy matrix. The monodromy matrix is a matrix of differentials that expresses how the time-evolved phase-space variables depend on the initial phase-space variables\cite{Irigoyen2006} 
\begin{equation}    
  \label{monodromy}
    \bo{M} \equiv \frac{\partial \bo{z}_t}{\partial \bo{z}_0}  \quad\text{where}\quad \e{M}_{\e{XY}} = \frac{\partial \e{X}(t)}{\partial \e{Y}(0)} \\ \vspace{+5pt}.
\end{equation}
The monodromy matrix is computed for each timestep and multiplied with the previous timestep matrices to obtain the monodromy matrix for the overall trajectory 
\begin{gather} 
 \label{monotimestep}
    \bo{M}(2 \Delta t) = \bo{M} (\Delta t \to \Delta 2t ) \bo{M}(0 \to \Delta t).
\end{gather}
If the Hamiltonian is split, $H = H_{1} + H_{2}$, then the monodromy matrix is calculated for each timestep according to the algorithmic flow map.\cite{Leimkuhler2005} Calculation of the monodromy matrix is not needed for algorithm propagation, only to determine symplecticity and if the theoretical framework requires it. For example, in some SC-IVR methods the phase-factor can be calculated using elements of the monodromy matrix, where utilising a symplectic algorithm can improve the stability, partially mitigating the notorious `sign' problem.\cite{Church2018, Malpathak2022, Venkataraman2007} 

The symplecticity criterion, Eqn.~\eqref{symplecticity}, is a much stricter condition than  the conservation of volume phase-space (Liouville’s theorem), which only requires the monodromy matrix determinant to be unity.\cite{Church2018, Hardy1995, Tuckerman2010} 
Volume phase-space preservation is a consequence of symplecticity, but this relationship does not necessarily hold the other way around,\cite{Leimkuhler2005} i.e. volume phase-space preservation is necessary but not sufficient for symplecticity. Instead, symplectic integrators can arise from exact time-propagation of a Hamiltonian system.\cite{Leimkuhler2005} Splitting the Hamiltonian into sub-evolutions will also result in a symplectic integrator, provided that each sub-evolution is the exact time-propagation of the relevant sub-Hamiltonian.\cite{Church2018, Leimkuhler2005} For example, velocity-Verlet is a classical symplectic algorithm as splitting results in two sub-Hamiltonians that are independent of each other, such that both can be integrated exactly.\cite{Hardy1995} The MMST Hamiltonian contains coupling between the electronic momenta and nuclear position, making symplectic time-evolution of the equations of motion challenging and in general, results in nonlinear dynamics.\cite{Church2018, Stock1997} Symplectic integrators are an advantage for Hamiltonian integration as they have little to no energy drift with time and tend to be more stable at long simulation times.\cite{Church2018} We also note that the Cayley transform can improve the symplectic stability of algorithms with no additional algorithmic complexity, computational cost and can be implemented for any path-integral based scheme.\cite{Korol2019}

\subsection{Algorithmic Overview}
The MInt algorithm, by Church \textit{et al.}~(2018), was developed to help extend the MQC-IVR method to simulate nonadiabatic dynamics using the MMST mapping.\cite{Church2018} The name arose as the MInt algorithm exactly solves the \textbf{M}omentum \textbf{Int}egral with time and as we will show, is the only known symplectic algorithm propagating the MMST Hamiltonian.\cite{Church2018} The MInt algorithm splits the MMST Hamiltonian into two sub-Hamiltonians, each of which is propagated exactly
\begin{subequations} \label{splitH} \begin{gather}  
    H = H_{1} +H_{2} \text{,} \\
    H_{1} = \frac{1}{2} \bo{p}^T \boldsymbol{\mu}^{-1} \bo{p} \text{,} \\
    H_{2} = \frac{1}{2} \left\{\bo{P}^T \bo{V}(\bo{x})\bo{P} + \bo{X}^T \bo{V}(\bo{x})\bo{X} - \mathrm{Tr}[\bo{V}(\bo{x})]\right\}. \end{gather}
\end{subequations} 
Exact evolution of the sub-Hamiltonians results in symplecticity, and the sub-evolution of $H_{1}$ is split into two half timesteps to improve the time-order error, such that the algorithm is at least a second-order method.\cite{Church2018, Leimkuhler2005} Hamilton’s equations of motion are obtained for $H_{1}$ and $H_{2}$, with the latter being more complicated due to coupling between the nuclear and quantum DoF.\cite{Church2018} The MInt algorithm can be used on any Hamiltonian containing a sum of Meyer--Miller-like terms and has algebraically been shown to be symplectic, symmetric, second-order in time and time-reversible.\cite{Church2018} Fortran code of the MInt algorithm is available in the SC-IVR package on the Ananth Group website.\cite{AnanthSoftware}
Recently, the MInt algorithm was utilised by Gardner \textit{et al.}~in \texttt{NQCDynamics.jl}, a Julia package for condensed phase nonadiabatic quantum dynamics.\cite{Gardner2022} To be able to compare the three algorithms on an equal theoretical footing, we use the D5 form of the MInt algorithm where instead $H_{2}$ is split into two half-timesteps\cite{Church2018}
\begin{gather}
    \label{MIntflow} \Psi_{H,\Delta t}^{\textrm{MInt}} := \Phi_{H_{2},\frac{\Delta t}{2}} \circ \Phi_{H_{1},\Delta t} \circ \Phi_{H_{2},\frac{\Delta t}{2}} \text{,}
\end{gather}
where $\Psi_{H,\Delta t}^{\textrm{MInt}}$ is the approximate flow map comprised of exact evolutions of the relevant sub-Hamiltonian, $\Phi_{H,\Delta t}$, and propagation is done from right to left. In our notation, $\Phi$ refers to exact evolution and $\Psi$ refers to approximate evolution which may or may not be comprised of exact sub-evolutions, consistent with the notation of Leimkulher and Reich.\cite{Leimkuhler2005}

The Split-Liouvillian (SL) algorithm, by Richardson \textit{et al.}~(2017), uses the Liouvillian formalism to construct an integrator.\cite{Richardson2017} The Liouvillian operator can be generated from the Hamiltonian using a Poisson bracket
\begin{gather} 
    \label{liouvillian}
    \mathscr{L} = -\{H, \cdot \}  \text{,}
\end{gather}
where the solutions to time-evolution are of exponential form.\cite{Tuckerman2010} We follow the convention in the Matsubara dynamics article\cite{Hele2015} and by Zwanzig,\cite{Zwanzig2001} and define the Liouvillian to be real with no factor $i$ (the imaginary unit).
The MInt algorithm D5 form, Eqn.~\eqref{MIntflow}, is equivalent in Liouvillian notation to\cite{Church2018} 
\begin{gather} \label{MIntliovillian}
    \Psi_{H, \Delta t}^{\textrm{MInt}} := e^{\mathscr{L}_{2} \frac{\Delta t}{2}} e^{\mathscr{L}_{1} \Delta t} e^{\mathscr{L}_{2} \frac{\Delta t}{2}} \text{,}
\end{gather}
where evolution requires propagation of the Liouvillians from the right to the left and $\mathscr{L}_{1/2}$ are the Liouvillians of $H_{1/2}$ respectively. The SL algorithm further symmetrically splits $\mathscr{L}_{2}$ into an electronic (\textrm{el}) and nuclear (\bo{p}) contribution. Hence, the flow map for this algorithm is
\begin{align} \label{SLflow}
    \Psi_{H, \Delta t}^{\textrm{SL}} &:= e^{\mathscr{L}_{\textrm{el}} \frac{\Delta t}{2}} e^{\mathscr{L}_{\bo{p}} \frac{\Delta t}{2}} e^{\mathscr{L}_{1} \Delta t} e^{\mathscr{L}_{\bo{p}} \frac{\Delta t}{2}}e^{\mathscr{L}_{\textrm{el}} \frac{\Delta t}{2}} \text{,} %\label{eqn14b}
\end{align}
where
\begin{subequations} \label{LelLp} \begin{align}
    \mathscr{L}_{\bo{p}} &=  \sum_{i} \dot{p_{i}} \cdot \frac{\partial}{\partial p_{i}} \text{,}\\
    \mathscr{L}_{\textrm{el}} &=  \sum_{i}\sum_{j} P_{ij}^\textrm{T}\textbf{V}(x_{i})\frac{\partial}{\partial X_{ij}} - X_{ij}^\textrm{T}\textbf{V}(x_{i})\frac{\partial}{\partial P_{ij}} \text{,} \end{align}
\end{subequations}
are the electronic and nuclear Liouvillians of $H_{2}$ respectively, where \textit{i} is the nuclear index and \textit{j} is electronic index.\cite{Church2018, Richardson2017} 
This uses the approximation
\begin{align} \label{slapprox}
    e^{\mathscr{L}_{2}\Delta t} =& \ e^{(\mathscr{L}_{\textrm{el}} +\mathscr{L}_{\bo{p}} )\Delta t} \simeq e^{\mathscr{L}_{\textrm{el}}\Delta t}  e^{\mathscr{L}_{\bo{p}}\Delta t} \text{,} \\
    e^{\mathscr{L}_{2}\Delta t} =& \ e^{(\mathscr{L}_{\bo{p}}+\mathscr{L}_{\textrm{el}})\Delta t} \simeq e^{\mathscr{L}_{\bo{p}}\Delta t} e^{\mathscr{L}_{\textrm{el}}\Delta t} \text{,} \nonumber
\end{align}
which is valid in the limit $\Delta t \to 0$.\cite{Richardson2017} Hence, the flow map can also be represented as
\begin{align} \label{SLflow2}
    \Psi_{H, \Delta t}^{\textrm{SL}} &:= \Psi_{\textrm{SL}^b, \frac{\Delta t}{2}} \circ \Phi_{H_{1}, \Delta t} \circ \Psi_{\textrm{SL}^a, \frac{\Delta t}{2}} \text{,}
\end{align}
where $\Psi_{\textrm{SL}^{a/b}}$ represents the approximate propagation of $H_{2}$. Due to symmetric splitting the two half timesteps are not equal and are labelled with the superscripts $a$ for $e^{\mathscr{L}_{\bo{p}}}  e^{\mathscr{L}_{\textrm{el}}}$ and $b$ for $e^{\mathscr{L}_{\textrm{el}}}  e^{\mathscr{L}_{\bo{p}}}$.
For an integrator to be symplectic, it is sufficient but not necessary that the integrator is a sequence of exact sub-Hamiltonian evolutions. Evolution under Liouvillians is not guaranteed to be symplectic but will be if each Liouvillian corresponds to exact propagation of a Hamiltonian or sub-Hamiltonian.\cite{Leimkuhler2005} For an arbitrary timestep, the splitting of $H_{2}$ into an electronic and nuclear contribution results in holding the electronic position and momenta constant while propagating the nuclear momenta. Hence, the propagation of $H_{2}$ is no longer exact or guaranteed to be symplectic.\cite{Church2018} Church \textit{et al.}~(2018) derived the SL monodromy matrix and using the symplecticity criterion, Eqn.~\eqref{symplecticity}, confirmed that the SL algorithm is not symplectic.\cite{Church2018} They stated that there is likely to be an energy drift associated but this may be small if the adiabatic states are close in energy and weakly coupled.\cite{Church2018} However, the algebraic proof does not indicate if the SL algorithm will result in unreasonable dynamics such as a very large energy drift when compared to the MInt for a given timestep.

Kelly \textit{et al.}~(2012) have also developed an integrator that we refer to as the DE algorithm due to the approximation needed to obtain the final algebraic form.  
The DE algorithm also splits the Hamiltonian into $H_{1}$ and $H_{2}$, where the motivation is to enhance stability and minimise the difference between the exact and approximate dynamics.\cite{Kelly2012} The algorithm uses MInt-like equations but makes an implicit approximation equivalent to assuming that the diabatic potential matrix eigenvalues are degenerate,\cite{Kim2014} simplifying the calculation of the potential matrix derivative, making the DE algorithm unlikely to be symplectic.\cite{Church2018, Kelly2012} The flow map cannot easily be expressed in a Liouvillian form but can be written as
\begin{gather}
    \label{DEflow} \Psi_{H,\Delta t}^{\textrm{DE}} := \Psi_{\textrm{DE
},\frac{\Delta t}{2}} \circ \Phi_{H_{1},\Delta t} \circ \Psi_{\textrm{DE},\frac{\Delta t}{2}} \text{,}
\end{gather}
where $\Psi_{\textrm{DE}}$ is the approximate DE propagation of $H_{2}$. The degenerate eigenvalue approximation has previously been utilised in PBME simulations of coherent dynamics in photosynthetic systems.\cite{Kelly2011} However, this is known to cause the state populations to deviate from the exact results for systems with large energy biases.\cite{Kim2014} 

The three algorithms use equivalent propagation equations for the nuclear position, electronic position and momenta, only differing in the nuclear momenta propagation.\cite{Kelly2012, Church2018, Richardson2017} The DE algorithm has a similar form to the SL algorithm, but the DE algorithm does not hold the electronic position and momenta stationary while propagating the nuclear momenta and has a different diabatic potential matrix differential with respect to nuclear position.\cite{Kelly2012, Richardson2017} As far as we are aware, the DE algorithm monodromy matrix has not been algebraically determined, tested for symplecticity or Liouville's theorem.

\subsection{\texorpdfstring{Propagation of $\bo{H_{1}}$}{Propagation of H1}}

For simplicity, we will consider only one nuclear DoF and two electronic DoF throughout as the generalised multi-dimensional form is known.\cite{Church2018, Richardson2017, Kelly2012} 
 All three algorithms propagate $H_{1}$ in the same way using Hamilton’s equations of motion\cite{Church2018, Richardson2017, Kelly2012} 
\begin{equation} \label{H1HEOM}
        \dot{\bo{X}} = 0 \quad\text{,}\quad  
        \dot{\bo{P}} = 0 \quad\text{,}\quad 
        \dot{x} = \frac{p}{m} \quad\text{,}\quad
        \dot{p} =  0 
\end{equation}
where $m$ is the nuclear mass, such that integration provides the propagation equations 
\begin{subequations} \label{H1HE} \begin{align}
        \bo{X}(t + \Delta t) =  \bo{X}(t) \quad &\text{,} \quad
        \bo{P}(t + \Delta t)   =   \bo{P}(t) \text{,} \\
        x(t + \Delta t)  =   x(t) + \frac{p(t)}{m} \Delta t \quad &\text{,} \quad
        p(t + \Delta t)   =  p(t) . 
        \end{align}
\end{subequations}
We define our monodromy matrix as
\begin{gather} \label{monoelementform}
    \bo{M} = \left[ {\begin{array}{cccc}
    \textrm{M}_{xx} & \bo{M}_{x\bo{X}} & \textrm{M}_{xp} & \bo{M}_{x\bo{P}}  \\
    \bo{M}_{\bo{X}x} & \bo{M}_{\bo{XX}} & \bo{M}_{\bo{X}p} & \bo{M}_{\bo{XP}}  \\
    \textrm{M}_{px} & \bo{M}_{p\bo{X}} & \textrm{M}_{pp} & \bo{M}_{p\bo{P}} \\
    \bo{M}_{\bo{P}x} & \bo{M}_{\bo{PX}} & \bo{M}_{\bo{P}p} & \bo{M}_{\bo{PP}}  \\
  \end{array} } \right] \text{,}
\end{gather}
such that the monodromy matrix for the propagation of $H_{1}$ is a triangular matrix of the form
\begin{equation}
    \label{H1Mono}
    \bo{M}_{H_1} = \left[ {\begin{array}{cccc}
    1 & \bo{0}^T & \frac{\Delta t}{M} & \bo{0}^T  \\
    \bo{0} & \mathbb{I} & \bo{0} & \mathbb{O}  \\
    0 & \bo{0}^T & 1 & \bo{0}^T  \\
    \bo{0} & \mathbb{O} & 0 & \mathbb{I}  \\
  \end{array} } \right] \text{,}
 \end{equation}
where $\bo{0} = [0,0]^\textrm{T}$ and the determinant is unity, $|\bo{M}_{H_{1}}| =1$. The propagation of $H_{1}$ is symplectic for all three algorithms as $\bo{M}_{H_{1}}^\textrm{T}\bo{J}^{-1}\bo{M}_{H_{1}} = \bo{J}^{-1}$.

\subsection{\texorpdfstring{Propagation of $\bo{H_{2}}$}{Propagation of H2}}
Here, we present and compare the propagation of ${H_{2}}$ for the three algorithms. The diabatic potential matrix is split into a state-independent term, $U$, and a traceless state-dependent matrix, $\tilde{\bo{V}}$, such that $\bo{V}(x) = U(x) +\tilde{\bo{V}}(x)$. This is as the DE algorithm requires a traceless state-dependent matrix, previously being seen as an advantage as it renders the dynamics invariant to any constant shift of the coupling potential.\cite{Kelly2012, Martinez2015} 
This gives
\begin{align} 
    \label{H2}\begin{split}
    H_{2} = U(x) + \frac{1}{2} \left\{ \bo{P}^T \tilde{\bo{V}}({x})\bo{P} + \bo{X}^T \tilde{\bo{V}}({x})\bo{X} - \mathrm{Tr}[\tilde{\bo{V}}({x})]\right\} \text{,}
    \end{split}
\end{align} 
where $\bo{X} = [X_{1}, \hdots, X_{n}]^\textrm{T}$ and similarly for \e{P}.
The SL and DE algorithms use approximations to propagate $H_{2}$ resulting in different, non-exact nuclear momentum propagation equations.\cite{Church2018, Richardson2017, Kelly2012}

The propagation of $H_{2}$ is split into two half-timesteps that sandwich $\Phi_{H_{1}}$, allowing fair algorithmic comparison with the original SL form and the equivalent MInt form. We have swapped the order of $H_{1}$ and $H_{2}$ to test the DE algorithm and in Figs.~5 and 6 in the SI, we show that this has no effect on the symplecticity or energy conservation. 

\subsubsection{The Momentum Integral (MInt) Algorithm}
The MInt algorithm exactly propagates $H_{2}$ with time, taking into account the electronic dependence of the nuclear momentum.\cite{Church2018} Using Hamilton's equations of motion for half a timestep, following the approach of Church \textit{et al.}~we derive propagation equations\begin{subequations} \label{H2MInt}
        \begin{align}
        &\dot{\bo{X}}  = \tilde{\bo{V}}\bo{P}  \quad \text{,} \quad
        \dot{\bo{P}} =  -\tilde{\bo{V}} \bo{X} \quad \text{,} \quad \dot{x} = 0 \text{,} \\\label{H2mP}
        &\dot{p} = -\ U' -\frac{1}{2}\left\{ (\bo{X} - i\bo{P})^\textrm{T}\tilde{\bo{V}}'(\bo{X}+ i\bo{P}) +\mathrm{Tr}[\tilde{\bo{V}}']\right\} \text{,}
        \end{align}
\end{subequations}
where the prime denotes the derivative with respect to \textit{x} and the matrix dependence on \textit{x} has been dropped for simplicity. To find solutions through integration, the MInt algorithm relies on the fact that $\dot{\bo{X}}$ and $\dot{\bo{P}}$ are independent of \textit{p} but $\dot{p}$ is dependant on \e{X} and \e{P}.\cite{Church2018} Therefore, solving for $\bo{X}(t+\frac{\Delta t}{2})$ and $\bo{P}(t+\frac{\Delta t}{2})$ and substituting into Eqn.~\eqref{H2mP} allows $p(t+\frac{\Delta t}{2})$ to be found. The electronic position and momenta are found through integration resulting in
\begin{align} \label{ELECTRONICPROP}
    &\left[\bo{X}\left(t+\frac{\Delta t}{2}\right) +i\bo{P}\left(t+\frac{\Delta t}{2}\right)\right]  = e^{-i \ \tilde{\bo{V}} \Delta t/2}\left[\bo{X}(t) +i\bo{P}(t)\right] \text{,} \end{align} 
which can be recast to be entirely real by diagonalizing $\tilde{\bo{V}}$ into eigenvectors, \e{S}, and a diagonal eigenvalue matrix, $\boldsymbol{\Lambda}$, such that $\bo{S}^\textrm{T}\tilde{\bo{V}}\bo{S} = \boldsymbol{\Lambda}$.\cite{Church2018} This results in
\begin{subequations} \label{electronicpropagationfinal} \begin{align}
         &\bo{X}\left(t+\frac{\Delta t}{2}\right) = \bo{C} \bo{X}(t) - \bo{D}\bo{P}(t) \text{,} \\&\bo{P}\left(t+\frac{\Delta t}{2}\right) = \bo{C} \bo{P}(t) + \bo{D} \bo{X}(t) \text{,} \end{align}
\end{subequations}
for the propagation of \e{X} and \e{P}, where $\bo{C}$ and $\bo{D}$ are,\cite{Church2018} 
\begin{align} \label{eqn24} &\bo{C} = \bo{S}\cos \left( \boldsymbol{\Lambda} \frac{\Delta t}{2} \right)\bo{S}^\textrm{T} \quad \text{,} \quad \bo{D} = \bo{S}\sin \left( -\boldsymbol{\Lambda} \frac{\Delta t}{2} \right)\bo{S}^\textrm{T} .
\end{align}
By defining the derivative of the potential in the adiabatic basis to be $\bo{G} = \bo{S}^\textrm{T}\tilde{\bo{V}}'\bo{S}$ and inserting $\bo{S}\bo{S}^\textrm{T} = \mathbb{I}$ identities into the integration of Eqn.~\eqref{H2mP}, an intermediate equation for the propagation of \textit{p} can be found
 \begin{align} \label{pintermediate} \begin{split}
    p\left(\frac{\Delta t}{2}\right) = p(0) - \frac{\Delta t}{2} U' -\frac{1}{2} \int_{0}^{\frac{\Delta t}{2}} \mathrm{d}t \ [\bo{X}(0) -i\bo{P}(0)]^\textrm{T} \\ \times \bo{S}e^{+i\boldsymbol{\Lambda}t}\bo{G}e^{-i\boldsymbol{\Lambda}t} \bo{S}^\textrm{T} [\bo{X}(0) +i\bo{P}(0)] \\{} + \frac{1}{2} \int_{0}^{\frac{\Delta t}{2}} \mathrm{d}t \, \mathrm{Tr}[\tilde{\bo{V}}'] \text{,} \end{split} \end{align}
which can be solved by element-wise integration of
\begin{align}
    \label{pcenter} \begin{split}
    \int_{0}^{\frac{\Delta t}{2}} \mathrm{d}t \ \bo{S}e^{+i\boldsymbol{\Lambda}t}\bo{G}e^{-i\boldsymbol{\Lambda}t}\bo{S}^\textrm{T} = \bo{E} + i\ \bo{F}. \end{split} 
\end{align} 
Defining
\begin{subequations} 
        \label{gammaxi} \begin{align}
        {\boldsymbol{\Gamma}}_{nm} &= \begin{cases} 
        {\boldsymbol{\Lambda}}'_{nn} \frac{\Delta t}{2}  &\ n=m\\
        -({\bf{S}}^\textrm{T}{\bf{S}}')_{nm}\sin (\lambda_{nm}\frac{\Delta t}{2} ) &\ n \neq m \text{,}
        \end{cases} \\
        {\boldsymbol{\Xi}}_{nm} &= \begin{cases} 
        0  &\ n=m\\
        [\cos(\lambda_{nm}\frac{\Delta t}{2} ) -1]({\bf{S}}^\textrm{T}{\bf{S}}')_{nm}&\ n \neq m \text{,}
        \end{cases} \end{align}
    \end{subequations}
where $\lambda_{nm} = ({\boldsymbol{\Lambda}})_{mm} - ({\boldsymbol{\Lambda}})_{nn} $, such that
\begin{subequations} 
        \label{EF} \begin{align}
        {\bf{E}} &:= {\bf{S}} {\boldsymbol{\Gamma}}  \bf{S}^\textrm{T} \text{,} \\
        {\bf{F}}  & :=  {\bf{S}} {\boldsymbol{\Xi}}  \bf{S}^\textrm{T} \text{,} \end{align}
\end{subequations}
where $\bf{E}$ is symmetric and $\bf{F}$ is skew-symmetric. The nuclear propagation, where $x$ is unchanged, is therefore
\begin{align} \label{mintpfinal}
    p\left(t + \frac{\Delta t}{2}\right)  \ &= \ \begin{aligned}[t] p(t)- {} \frac{\Delta t}{2} U' -\frac{1}{2} \left[ \bo{X}^\textrm{T}(t)\bo{E}\bo{X}(t) \right. \\ \left.+ {}\bo{P}^\textrm{T}(t)\bo{E}\bo{P}(t) -2\bo{X}^\textrm{T}(t)\bo{F}\bo{P}(t) \right]  \\ {}+ \frac{\Delta t}{4}\mathrm{Tr}[\tilde{\bo{V}}']. \end{aligned}
\end{align}
The extended form for multiple states can be found in Appendix B of Ref.~[\!\!\citenum{Church2018}].
The MInt monodromy matrix for the propagation of $H_{2}$ satisfies Liouville's theorem and can be found by defining
\begin{subequations} \label{monoelements} \begin{align}
    \bo{a} &= -\bo{P}^\textrm{T}\bo{E} + \bo{X}^\textrm{T}\bo{F} \text{,} \\
    b &= \begin{aligned}[t] -\frac{\Delta t}{2} U'' {} -\frac{1}{2}\left( \bo{X}^\textrm{T}\bo{E}'\bo{X} + \bo{P}^\textrm{T}\bo{E}'\bo{P} - 2\bo{X}^\textrm{T}\bo{F}'\bo{P}\right) \\ + {} \frac{1}{4} \mathrm{Tr}[\bo{\tilde{V}}'']\Delta t \text{,} \end{aligned} \\
    \bo{e} &= -\bo{X}^\textrm{T}\bo{E} -\bo{P}^\textrm{T}\bo{F} \text{,}\\
    \bo{f} &= \bo{C}'\bo{P} +\bo{D}'\bo{X} \text{,}\\
    \bo{g} &= \bo{C}'\bo{X} - \bo{D}'\bo{P} \text{,}
    \end{align}
\end{subequations}
such that the monodromy matrix is\cite{Church2018} 
\begin{equation}
    \label{H2monomint}
    \bo{M}_{H_2} = \left[ {\begin{array}{cccc}
    1 & \pmb{0}^\textrm{T}  & 0 & \pmb{0}^\textrm{T}  \\
    \bo{g}  & \bo{C} & \pmb{0} & -\bo{D}  \\
    b & \bo{e} & 1 & \bo{a} \\
    \bo{f} & \bo{D} & \pmb{0} & \bo{C}  \\
  \end{array} } \right] \quad \text{and}\quad |\bo{M}_{H_{2}}| = 1.
 \end{equation}
Propagation under $H_{2}$ is symplectic as $\bo{M}_{H_{2}}$ satisfies the symplecticity criterion, Eqn.~\eqref{symplecticity}, derived by Church \textit{et al.}~and shown in Appendix \ref{symp}. As propagation under both $H_{1}$ and $H_{2}$ is symplectic, the overall propagation for the MInt algorithm is symplectic.\cite{Church2018, Leimkuhler2005}

\subsubsection{The Split-Liouvillian (SL) Algorithm}
The Split-Liouvillian (SL) algorithm propagates $H_{2}$ using the Liouvillian formalism by splitting further into an electronic and nuclear momentum propagation, where Eqn.~\eqref{LelLp} becomes\cite{Richardson2017, Church2018}
\begin{align} \label{LelLpfinal}
    &\mathscr{L}_{\textrm{el}} =  \bo{P}^\textrm{T}\tilde{\bo{V}}\nabla_{\bo{X}} - \bo{X}^\textrm{T}\tilde{\bo{V}}\nabla_{\bo{P}} \text{,}\\ 
    &\mathscr{L}_{\bo{p}}  = - \left\{ U' + \frac{1}{2}[(\bo{X}- i\bo{P})^\textrm{T}\tilde{\bo{V}}'(\bo{X} + i\bo{P}) -\mathrm{Tr}[\tilde{\bo{V}}']] \right\} \frac{\partial}{\partial p}. \nonumber
\end{align}
The electronic propagation resulting from $\mathscr{L}_{\textrm{el}}$ is equivalent to the MInt algorithm, Eqn.~\eqref{electronicpropagationfinal}, and leaves the nuclear variables unchanged.\cite{Church2018}
$\mathscr{L}_{\bo{p}}$ results in the following propagation of \textit{p} and leaves all other variables unchanged
\begin{align} \label{pslprop}
   p\left(t + \frac{\Delta t}{2}\right)  \ = \ & p(t) -\frac{\Delta t}{2} U' -\frac{\Delta t}{4} \left\{ \bo{X}^\textrm{T}\left( \frac{\Delta t}{2}\right) \tilde{\bo{V}}'\bo{X} \left( \frac{\Delta t}{2}\right) \right. \nonumber \\  &+ \left. \bo{P}^\textrm{T}\left( \frac{\Delta t}{2} \right) \tilde{\bo{V}}'\bo{P} \left( \frac{\Delta t}{2} \right) -\mathrm{Tr}[\tilde{\bo{V}}'] \right\} . \end{align}
The difference between the two algorithms arises in the nuclear momentum, where due to the SL symmetric propagation of $H_{2}$,  
the electronic variables are always the half-timestep evolved values for propagation of \textit{p}.\cite{Richardson2017, Church2018} 

The SL monodromy matrices for the propagation of $H_{2}$ obey Liouville's theorem and are
\begin{align} \label{H2monosl} \begin{split}
    \bo{M}_{\bo{p}} &= \left[ {\begin{array}{cccc}
    1 & \pmb{0}^\textrm{T}  & 0 & \pmb{0}^\textrm{T}  \\
    \pmb{0}  & \mathbb{I} & \pmb{0} & \mathbb{O}  \\
    \tilde{b} & - \frac{\Delta t}{2}\bo{X}^\textrm{T}\tilde{\bo{V}}' & 1 & - \frac{\Delta t}{2}\bo{P}^\textrm{T}\tilde{\bo{V}}'  \\
    \pmb{0} & \mathbb{O} & \pmb{0} & \mathbb{I}  \\
  \end{array} } \right]  \text{,} \\ %\\ \label{eqn48}
  \bo{M}_{\textrm{el}} &= \left[ {\begin{array}{cccc}
    1 & \pmb{0}^\textrm{T}  & 0 & \pmb{0}^\textrm{T}  \\
    \bo{g}  & \bo{C} & \pmb{0} & -\bo{D}  \\
    0 & \bo{0}^\textrm{T} & 1 & \pmb{0}^\textrm{T}  \\
    \bo{f} & \bo{D} & \pmb{0} & \bo{C}  \\
  \end{array} } \right] \text{,} \end{split}
\end{align}
where\cite{Church2018}
\begin{equation*}\label{slb}
    \tilde{b} = -\frac{\Delta t}{2} U'' -\frac{\Delta t}{4}\left( \bo{X}^\textrm{T}\tilde{\bo{V}}''\bo{X} + \bo{P}^\textrm{T}\tilde{\bo{V}}''\bo{P} - \mathrm{Tr}[\tilde{\bo{V}}'']\right). 
\end{equation*}
Neither $\bo{M}_{\bo{p}}$ and $\bo{M}_{\textrm{el}}$ satisfy the symplecticity criterion individually, seen in Appendix \ref{symp}.\cite{Church2018} The combination of $\bo{M}_{\bo{p}}$ and $\bo{M}_{\textrm{el}}$ was shown by Church \textit{et al.}~to only be symplectic in the $\Delta t \to 0$ limit, not for an arbitrary timestep.

\subsubsection{The Degenerate Eigenvalue (DE) Algorithm}

The DE algorithm defines $\tilde{\bo{V}}$ to be traceless, such that the last term of $H_{2}$ is ignored.\cite{Kelly2012} Here, we will re-frame the algorithm into a MInt-like form using vector notation and outline the degenerate eigenvalue approximation used. We will also investigate the symplecticity and satisfaction of Liouville's theorem through defining the monodromy matrix, which as far as we are aware has not been done before.

The electronic Hamiltonian equations are the same as above for the MInt and SL algorithms and solved to provide the same overall propagation equations, Eqn.~\eqref{electronicpropagationfinal}. The nuclear momentum propagation equation is
\begin{align} \label{DEpprop}
    \dot{p} =  -U' - \frac{1}{2}(\bo{X}^\textrm{T} \tilde{\bo{V}}'\bo{X} + \bo{P}^\textrm{T}\tilde{\bo{V}}'\bo{P}) .
\end{align}
To solve Eqn.~\eqref{DEpprop}, we define an integral, \textit{A}, that requires the degenerate eigenvalue assumption to arrive at the final form by Kelly \textit{et al.}~(2012)
\begin{align} \label{pintegralsl} 
    p \left( \frac{\Delta t}{2} \right) = \  p(0) - \frac{\Delta t}{2}U'  -  \frac{1}{2} \int_{0}^{\frac{\Delta t}{2}} \mathrm{d} \tau &\left[ \bo{X}^\textrm{T} (\tau) \tilde{\bo{V}}'\bo{X}(\tau)  \right. \\ &+ \bo{P}^\textrm{T}( \tau)\tilde{\bo{V}}'\bo{P}( \tau) \left. \right]  \nonumber \\
     =: \ p(0) - \frac{\Delta t}{2}U' +A \text{.}
\end{align}
This can be written in a MInt-like form
\begin{align} \label{Aintegral1} \begin{split}
    A = -\frac{1}{2} \int_{0}^{\frac{\Delta t}{2}} \mathrm{d} \tau \ [\bo{X}(\tau) - i\bo{P}(\tau)]^\textrm{T} \ \tilde{\bo{V}}'\ [\bo{X}(\tau) + i\bo{P}(\tau)] \textrm{,} 
    \end{split}
\end{align}
such that using Eqn.~\eqref{ELECTRONICPROP} 
\begin{align} \label{Aintegral2} \begin{split}
    A = -\frac{1}{2} \int_{0}^{\frac{\Delta t}{2}} \mathrm{d} \tau \ [\bo{X}(0) - i\bo{P}(0)]^\textrm{T}e^{i \ \tilde{\bo{V}} \tau /\hbar} \ \tilde{\bo{V}}'\  e^{-i\ \tilde{\bo{V}} \tau /\hbar} \\ \times[\bo{X}(0) + i\bo{P}(0)] \textrm{.} 
    \end{split}
\end{align}
A transformation is defined into the adiabatic basis using the following overlined variables\cite{Kelly2012} 
\begin{gather} 
\label{adiabatictransform}
   \overline{\bo{X}} = {\bo{S}}^{\textrm{T}} \bo{X} \quad\text{,}\quad   \overline{\bo{P}} = {\bo{S}}^{\textrm{T}} \bo{P}  \text{,}
\end{gather}
such that, conversion into the adiabatic basis utilising the decomposition of $\tilde{\bo{V}}$ results in
\begin{gather} \label{Aintegral3} \begin{split}
    A = -\frac{1}{2} \int_{0}^{\frac{\Delta t}{2}} \mathrm{d} \tau &\quad [\overline{\bo{X}}(0) - i\ \overline{\bo{P}}(0)]^\textrm{T} e^{i\boldsymbol{\Lambda}\tau/\hbar}\bo{S}^\textrm{T}
    \tilde{\bo{V}}' \bo{S} \\ &\times e^{-i\boldsymbol{\Lambda}\tau/\hbar}[\overline{\bo{X}}(0) + i\ \overline{\bo{P}}(0)] \text{,} 
    \end{split}
\end{gather}
where $\tilde{\bo{V}}'$ is determined by
\begin{gather}\label{adiabaticV} \begin{split}
    &\tilde{\bo{V}}' = \bo{S}' \boldsymbol{\Lambda} \bo{S}^\textrm{T} + \bo{S} \boldsymbol{\Lambda}' \bo{S}^\textrm{T} +\bo{S} \boldsymbol{\Lambda} {\bo{S}^{\textrm{T}}}' \text{,} \\\text{such that}\quad %\label{eqn64}
     &\bo{S}^\textrm{T}\tilde{\bo{V}}' \bo{S} = \bo{S}^\textrm{T} \bo{S}' \boldsymbol{\Lambda} +  \boldsymbol{\Lambda}' +\boldsymbol{\Lambda} {\bo{S}^{\textrm{T}}}' \bo{S} .\end{split}
\end{gather}
%\vspace{-2pt}
We can arrive at the form shown in Ref.~[\!\!\citenum{Kelly2012}] by making the approximation that the eigenvalues are equal, $\boldsymbol{\Lambda} \simeq \epsilon\mathbb{I}$. Therefore, the differential of the eigenvalues can be approximated as $\pmb{\Lambda}' \simeq \epsilon'\mathbb{I}$ such that the derivative of $\tilde{\bo{V}}'$ in the adiabatic basis is obtained through differentiating the identity, $ \mathbb{I}' = (\bo{S}\bo{S}^{\textrm{T}})' = \bo{S}^\textrm{T} \bo{S}' + {\bo{S}^{\textrm{T}}}' \bo{S}=0$,
\begin{align} \label{Gbasis}
    {\bo{G}} &= \bo{S}^\textrm{T}\tilde{\bo{V}}' \bo{S} = \bo{S}^\textrm{T} \bo{S}' \boldsymbol{\Lambda} +  \boldsymbol{\Lambda}' +\boldsymbol{\Lambda} {\bo{S}^{\textrm{T}}}' \bo{S}
 \nonumber \\ &\simeq  \pmb{\Lambda}' + \epsilon(\bo{S}^\textrm{T} \bo{S}' + {\bo{S}^{\textrm{T}}}' \bo{S}) \ = \ \boldsymbol{\Lambda}'.
\end{align}
Inserting this into Eqn.~\eqref{Aintegral3} results in
\begin{gather} \label{innerintegral}
      e^{i\pmb{\Lambda}\tau/\hbar} \pmb{\Lambda}'e^{-i\pmb{\Lambda}\tau/\hbar} \simeq  \epsilon' e^{i\pmb{\Lambda}\tau/\hbar} \mathbb{I}e^{-i\pmb{\Lambda}\tau/\hbar} =  \epsilon' \mathbb{I} \simeq \pmb{\Lambda}'.
\end{gather}
Hence, the integral below is obtained
\begin{gather} \label{Afinal} \begin{split}
     A &\simeq -\frac{1}{2} \int_{0}^{\frac{\Delta t}{2}} \mathrm{d} \tau \ [\overline{\bo{X}}(0) - i\ \overline{\bo{P}}(0)]^\textrm{T} \pmb{\Lambda}'[\overline{\bo{X}}(0) + i\ \overline{\bo{P}}(0)] \\ 
    &= -\frac{\Delta t}{4} \left[ \overline{\bo{X}}^\textrm{T}(0)\boldsymbol{\Lambda}'\ \overline{\bo{X}}(0) +\overline{\bo{P}}^\textrm{T}(0)\pmb{\Lambda}'\ \overline{\bo{P}}(0) \right] \text{.} 
    \end{split}
\end{gather}
The final form seen in Ref.~[\!\!\citenum{Kelly2012}] has an additional $-\frac{\Delta t}{4}\mathrm{Tr}[\pmb{\Lambda}']$ term when written in matrix form, however, this is just a dummy term. The fact that $\tilde{\bo{V}}$ is traceless requires that the sum of the eigenvalues is zero, therefore, $\pmb{\Lambda}$ and $\pmb{\Lambda}'$ are traceless. The overall nuclear propagation in the diabatic basis is then 
\begin{align} \label{pfinalDE} 
         p\left(t + \frac{\Delta t}{2}\right) 	&= p(t) - \frac{\Delta t}{2}U' \begin{aligned}[t] -\frac{\Delta t}{4} \left[ \bo{X}^\textrm{T}(t) \ \bo{S}\boldsymbol{\Lambda}'\bo{S}^\textrm{T} \ {\bo{X}}(t) \right. \\ {}  \left. +\bo{P}^\textrm{T}(t) \ \bo{S}\boldsymbol{\Lambda}'\bo{S}^\textrm{T} \ \bo{P}(t) \right] .\end{aligned}
\end{align}
To obtain the final form by Kelly \textit{et al.}~(2012), we make the DE approximation to simplify Eqn.~\eqref{Gbasis}.\cite{Kelly2012, Kim2014} 

Comparison with the SL and MInt algorithms can determine whether the DE algorithm is likely to be symplectic, where the propagation of \textit{x}, \e{X} and \e{P} are equivalent.\cite{Kelly2012, Richardson2017, Church2018} It can be seen that in the case where the DE approximation holds such that, $\tilde{\bo{V}}'= \bo{S}\boldsymbol{\Lambda}'\bo{S}^\textrm{T}$, the propagation of \textit{p} would be similar to the SL algorithm. However, the DE algorithm uses the inital electronic variables when the SL algorithm uses the half-timestep evolved values.  
Due to the similarities prior to the DE approximation being made, the DE algorithm is unlikely to be symplectic. To rigorously check the symplecticity, we derive the monodromy matrix by defining
\begin{gather*} \label{Nde} 
    \bo{N} = (\bo{S}\pmb{\Lambda}'\bo{S}^\textrm{T})' = \bo{S}'\pmb{\Lambda}'\bo{S}^\textrm{T} +\bo{S}\pmb{\Lambda}''\bo{S}^\textrm{T} +\bo{S}\pmb{\Lambda}'\bo{S}'^{\textrm{T}} \text{,} \\ \label{bDE} \overline{b} = -\frac{\Delta t}{2}U'' - \frac{\Delta t}{4}(\bo{X}^\textrm{T} \bo{N} \bo{X} + \bo{P}^\textrm{T} \bo{N} \bo{P} ) \text{,}
\end{gather*}
such that
\begin{equation}
    \label{Mde}
    \bo{M}_{\textrm{DE}} = \left[ {\begin{array}{cccc}
    1 & \pmb{0}^\textrm{T}  & 0 & \pmb{0}^\textrm{T}  \\
    \bo{g}  & \bo{C} & \pmb{0} & -\bo{D}  \\
    \overline{b} & - \frac{\Delta t}{2}\bo{S}\pmb{\Lambda}'\bo{S}^\textrm{T}\bo{X} & 1 & - \frac{\Delta t}{2}\bo{S}\pmb{\Lambda}'\bo{S}^\textrm{T}\bo{P}  \\
    \bo{f}  & \bo{D} & \pmb{0} & \bo{C}  \\
  \end{array} } \right] .
 \end{equation}
The determinant of $\bo{M}_{\textrm{DE}}$ can easily be shown to be unity, satisfying Liouville's theorem.
However, the approximate propagation of $H_{2}$ is not symplectic as $\bo{M}_{\textrm{DE}}$ does not satisfy the symplecticity criterion, shown in Appendix \ref{symp}. The MInt algorithm is the only algorithm considered here that is both symplectic and satisfies Liouville's theorem.\cite{Church2018}

Below, we present a table of the theoretical results for easy algorithmic comparison.
\vspace{-5pt}
\begin{center}
    \begin{tabular}{p{0.36\linewidth}|p{0.11\linewidth}|p{0.14\linewidth}|p{0.18\linewidth}}
      \toprule % <-- Toprule here
      \textbf{Theoretical Results} & \textbf{ MInt } & \textbf{ SL } & \textbf{ DE }\\
      \midrule % <-- Midrule here
      \rowcolor[HTML]{C0C0C0}
        $H_1$ Propagation & Exact & Exact & Exact \\
      $H_{2}$ $\bo{X}/\bo{P}$ Propagation & Exact & Exact & Exact \\
      \rowcolor[HTML]{C0C0C0} $H_{2}$ $p$ Propagation & Exact & Approx. & Approx. \\
      $H_2$ Approximation & None & Split into $\mathscr{L}_{\textrm{el}}$ and $\mathscr{L}_{\bo{p}}$ & Approx. Degenerate Eigenvalues of $\bo{V}(\bo{x})$ \\
      \rowcolor[HTML]{C0C0C0}
      Satisfies Liouville's Theorem & \cmark & \cmark & \cmark\\
      Symplectic & \cmark & \xmark & \xmark \\
      \rowcolor[HTML]{C0C0C0}
      Exact in $\Delta t \to 0$ limit & \cmark & \cmark & \xmark \\
      
      \bottomrule % <-- Bottomrule here
    \end{tabular}
    \vspace{-5pt}
    \captionof{table}[Summary of the theoretical results obtained in here and Ref.~1. The $H_{2}$ electronic propagation for all algorithms is equivalent. The approximations made in the SL and DE algorithms affect the nuclear momentum propagation only.]{Summary of the theoretical results obtained here and Ref.~[\!\!\citenum{Church2018}].  The $H_{2}$ electronic propagation for all algorithms $H_{2}$ is equivalent. The approximations made in the SL and DE algorithms result in inexact nuclear momentum propagation. However, the SL algorithm is exact in the $\Delta t \to 0$ limit. }
    \label{theoreticaltable}
\end{center}

\section{Results and Discussion} \label{results}
The algorithms discussed here have been utilised in the literature but not for the same system to allow direct comparison. We seek to test the algorithms computationally on an equal footing using the same system. Hence, we use a simple two-state linear vibronic potential, also known as a double well potential, discussed in this section with the MMST Hamiltonian (corresponding to the single-bead limit of NRPMD) to test the symplecticity, satisfaction of Liouville's theorem, energy conservation and accuracy of correlation functions. The models defined here also allow qualitative comparison with literature.
\subsection{Theoretical Models}
To compare the approximate dynamics produced by the MInt, SL and DE algorithms, we use the three potential models introduced in Ref.~[\!\!\citenum{Richardson2013a}]. 
For these models the potential diabatic matrix is 
\begin{gather} \label{potentialmatrix}
    \bo{V} = \left[ {\begin{array}{cc} \frac{1}{2}m\omega^2x^2 +\alpha +\kappa x & \Delta \\
    \Delta & \frac{1}{2}m\omega^2x^2 -\alpha - \kappa x \end{array}} \right] \text{,}
\end{gather}
such that splitting to obtain a traceless $\tilde{V}$ gives
\begin{align} \label{splitting}
    %\bo{V} =&  \quad \bo{U}+\tilde{\bo{V}} \\ 
    U= \frac{1}{2}m\omega^2x^2 \quad , \quad \tilde{\bo{V}} = \left[ {\begin{array}{cc}  \alpha +\kappa x & \Delta \\
    \Delta & -\alpha - \kappa x \end{array}} \right].
\end{align} 
Reduced units are used where $ m = \hbar = \omega = 1$, so energy is measured in units of the frequency, $\omega$. The models represent bound potentials, defined in Table \ref{table1}, where $\Delta$ is the electronic coupling, $2\alpha$ is the energy bias between the potential energy surfaces (the asymmetry) and $\kappa$ is the vibronic coupling, chosen to be 1. Model 1 represents strong electronic coupling, where the nuclear dynamics occur on a longer timescale than the electronic oscillations and nuclear motion occurs in a mean field of the diabatic surfaces. Model 2 has a strong energy bias and the system is in the inverted Marcus regime. Model 3 is a challenging intermediate regime in which the timescales of the electronic and nuclear dynamics are similar.
\begin{center}
    \begin{tabular}{l|c|c|c}
      \toprule % <-- Toprule here
      \textbf{Model} & \textbf{$\quad \alpha \quad$} & \textbf{$\quad \Delta\quad$} & \textbf{Regime}\\
      \midrule % <-- Midrule here
     \rowcolor[HTML]{C0C0C0} 1 & 0 & 4 & Adiabatic Limit\\
      2 & 2 & 1 & Inverted Marcus Regime\\
     \rowcolor[HTML]{C0C0C0} 3 & 0 & 1 & Intermediate Regime\\
      \bottomrule % <-- Bottomrule here
    \end{tabular}
    \captionof{table}{Values for the potential matrix constants for Models 1-3 from Ref.~[\!\!\citenum{Richardson2013a}].} \label{table1}
    %\vspace{-25pt}
\end{center}

Models 1--3 were utilised to compute correlation functions using the following distribution, $\rho$, in the $N$-bead form in Refs.~[\!\!\citenum{Richardson2013a}, \citenum{Richardson2017}].
Here, for simplicity, we use the single-bead form 
\begin{align}
    \label{distribution}
    \rho =  \frac{4}{\pi^2} e^{-|{\bo{X}^2}| -|{\bo{P}^2|} -\beta(U + p^2/2m)} \text{,}
    %\times |W|
\end{align}
such that the partition function is 
\begin{align}
    \label{pfW}
    Z = \langle W \rangle_{\rho} \text{,}
\end{align}
where $W = {\bf{P}}^\textrm{T}{\bf{M}}{\bf{X}\bf{X}}^\textrm{T}{\bf{M}} {\bf{P}}$ and ${\bf{M}}=e^{-\beta{\tilde{\bf{V}}}/2}$.
Note that $W$ is positive definite for the single bead case. Monte Carlo importance sampling is utilised as detailed in Appendix \ref{samp}.
Time-independent equilibrium properties are calculated, for example, the population of state $n$\cite{Richardson2013a, Richardson2017} 
\begin{align}
    \label{popn}
    \frac{1}{Z}\text{tr}\left[e^{-\beta\hat{H}}\ket{n}\bra{n}\right] \simeq \frac{\langle A_{n} \ W \rangle_{\rho}}{\langle W \rangle_{\rho}} \text{.}
    %\frac{1}{Z}\text{tr}\left[e^{-\beta\hat{H}}\ket{n}\bra{n}\right] = \frac{\langle A_{n} \ \text{sgn}(W) \rangle_{\rho}}{\langle \text{sgn}(W) \rangle_{\rho}}
\end{align}
The derivation is similar to that of Eqn.~\eqref{partition1} in Appendix \ref{samp} by summing over all indices except $n$,\cite{Richardson2017, Richardson2013a} resulting in
\begin{align}
    \label{popn2} \begin{split}
    \frac{1}{Z}\text{tr}\left[e^{-\beta\hat{H}}\ket{n}\bra{n}\right] \simeq \frac{1}{Z}\frac{1}{2\pi\hbar}\int_{-\infty}^{\infty} \rho W \frac{{P}_{n}[{\bf{M}}{\bf{X}}]_n}{\bo{P}^\textrm{T}{\bf{M}}\bo{X}} \\ \times \ \mathrm{d}x\mathrm{d}p\mathrm{d}{\bf{X}}\mathrm{d}{\bf{P}} \text{,} \end{split} 
\end{align}
where,
\begin{align}
    \label{An}
    A_{n} &=  \frac{P_{n}[{\bf{MX}}]_{n}}{\bo{P}^\textrm{T}\bo{MX}} \text{.}
\end{align}
This approach for obtaining population information is only valid at $t=0$, whereas the electronic populations can be found at time $t$ using\cite{Richardson2017}
\begin{align}
    \label{Bn}
    B_{n} &= \frac{1}{2}\left( X_{n}^2 + P_{n}^2 -1\right).
\end{align}
Correlation functions are calculated through finding an approximation to Eqn.~\eqref{qtcf}, where for the position auto-correlation function $\hat{A} = \hat{B} = x$ and for the population auto-correlation function $\hat{A} = A_n$ and $ \hat{B} = B_n$,
\begin{subequations}
    \label{cf} \begin{align}
    \tilde{C}_{xx}(t) &\simeq  \frac{\langle x(0) \ x(t) \ W \rangle_{\rho}}{\langle W \rangle_{\rho}} \text{,} \\ \tilde{C}_{nn}(t) &\simeq \frac{\langle A_{n}(0) \ B_{n}(t) \ W \rangle_{\rho}}{\langle W \rangle_{\rho}} \text{,} \end{align}
\end{subequations}
where $W({\bf{X}}_{i}(0),{\bf{P}}_{i}(0))$ and initial conditions are sampled from Eqn.~\eqref{sampleddistribution} for $J$ trajectories.
The correlation functions are then averaged over $J$ trajectories
\begin{subequations}
    \label{cfsamp}
    \begin{align}
    \tilde{C}_{xx}(t) &= \frac{\sum_{i=1}^J x_{i}(0)\ x_{i}(t) \ W_{i}(0)}{ \sum_{i=1}^J W_{i}(0) } \text{,} \\ \tilde{C}_{nn}(t) &= \frac{\sum_{i=1}^J A_{ni}(0)\ B_{ni}(t) \ W_{i}(0)}{ \sum_{i=1}^J W_{i}(0) } \text{,}
    \end{align}
\end{subequations}
where index $i$ refers to the $i$th trajectory.
We can directly compare the three algorithms and qualitatively compare our results with Ref.~[\!\!\citenum{Richardson2013a}]. We sample the same distribution, with our method outlined in Appendix \ref{samp}, as in Ref.~[\!\!\citenum{Richardson2013a}]. However, we choose to split the potential matrix such that $\tilde{\bo{V}}$ is traceless as the DE algorithm requires this. In Ref.~[\!\!\citenum{Richardson2013a}], the matrix was instead split such that the lowest eigenvalue of 
 $\tilde{\bo{V}}$ is zero, which in general results in a non-zero trace. To ensure we can compare all three algorithms for the same Hamiltonian, we have used the traceless form leading to small quantitative differences between the results here and in Ref.~[\!\!\citenum{Richardson2013a}].

\subsection{Algorithmic Properties}
First, we consider the symplecticity and conservation of Liouville's theorem using Model 1. Church \textit{et al.}~determined that the MInt algorithm is symplectic whilst the SL algorithm is not.\cite{Church2018} In Appendix \ref{symp}, we algebraically show that the DE algorithm is not symplectic. To numerically determine the symplecticity, we define an error matrix, $\bo{E_{r}}$, to be
\begin{align} \label{errormatrix}
    \bo{E_{r}} &= \bo{M}^\textrm{T}\bo{J}^{-1}\bo{M} \ -  \ \bo{J}^{-1} \text{,}
\end{align}
where for a symplectic integrator the elements of $\bo{E_{r}}$, $a_{ij}$, will all be zero.\cite{Ivanov2013}
The Frobenius Norm is used to track the size of $\bo{E_{r}}$
\begin{align} 
    \label{frobnorm}
    ||{\bo{E_{r}}}||_{F} &= \sqrt{\sum_{i=1}^n\sum_{j=1}^n |a_{ij}|^2} \text{,}
\end{align}
where the matrix size is $n \times n$.\cite{Ivanov2013}
To average over many trajectories, we weight by $W$ 
\begin{align} 
    \label{frobnormweighted}
    \langle ||{\bo{E_{r}}}||_{F} \rangle_{\rho} &= \frac{\sum_{i=1}^J (||{\bo{E_{r}}}||_{F})_{i}(t) W_{i}(0)}{\sum_{i=1}^J 
 W_{i}(0)} \text{,}
\end{align}
where $i$ refers to the trajectory index.
To determine if Liouville's theorem is satisfied, we evaluate
\begin{align} 
    \label{determinantweighted}
    \langle (|{\bf{M}}|-1)^2 \rangle_{\rho} &= \frac{\sum_{i=1}^J ((|{\bf{M}}_{i}(t)|-1)^2 W_{i}(0)}{\sum_{i=1}^J 
 W_{i}(0)} \text{,}
\end{align}
which will be zero if it is satisfied.\cite{Church2018, Hardy1995} By squaring the deviation, we ensure no error cancellation when averaging over trajectories.

\begin{figure}[H]
\centering
\includegraphics[width=0.49\textwidth]{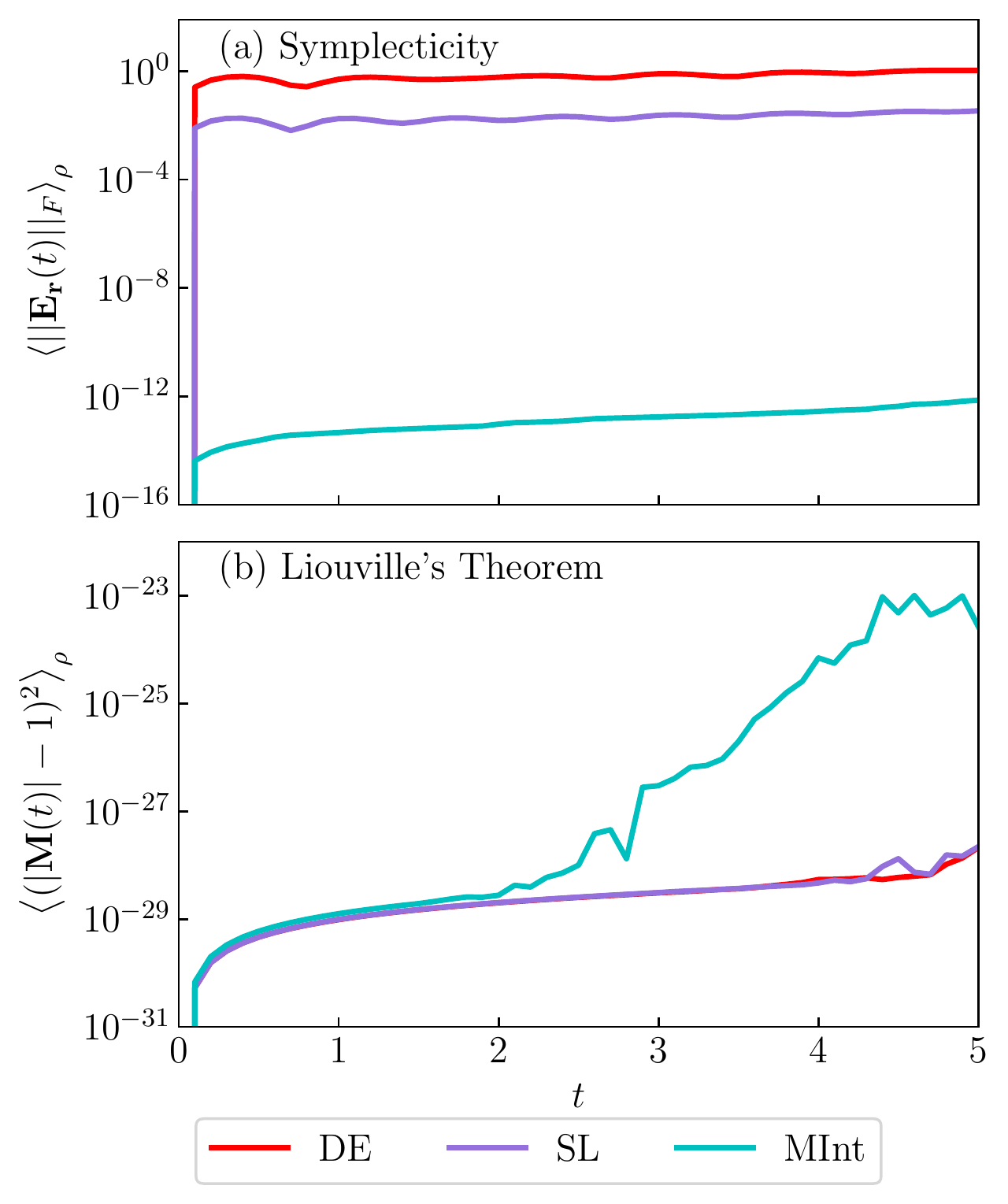}
\caption{(a) The Frobenius Norm of the symplecticity error matrix and (b) the determinant criterion as a function of time using Model 1 and $\Delta t = 0.1$, averaged over a million trajectories using the SL (purple), the MInt (cyan) and the DE (red) algorithms. In (a) the MInt algorithm is seen to be symplectic whereas the DE and SL are not, whereas in (b) all algorithms satisfy Liouville's theorem.}
\label{fig1}
\vspace{-10pt}
\end{figure}

In Fig.~\ref{fig1}a, the logarithmic plot of $||{\bo{E_{r}}}||_{F}$ against time for Model 1 with $\Delta t =0.1$ can be seen. The MInt algorithm (cyan) remains below $10^{-12}$ for the entire simulation time and is therefore symplectic, with a small build up of floating-point errors that arise in numerical calculations, in agreement with the literature.\cite{Church2018} The SL algorithm (purple) increases rapidly to around $10^{-2}$ and continues to increase indicating that it is not symplectic. The DE algorithm (red) is the least symplectic, being on the order of 1 by the end of the simulation time. Although the SL and DE algorithms are not symplectic, all three algorithms satisfy Liouville's theorem and conserve volume phase-space, as seen in Fig.~\ref{fig1}b. We believe that the very slight increase seen for the MInt algorithm is due to additional floating-point error accumulation arising from the more complicated propagation equations.

\begin{figure}[H]
\centering
\includegraphics[width=0.49\textwidth]{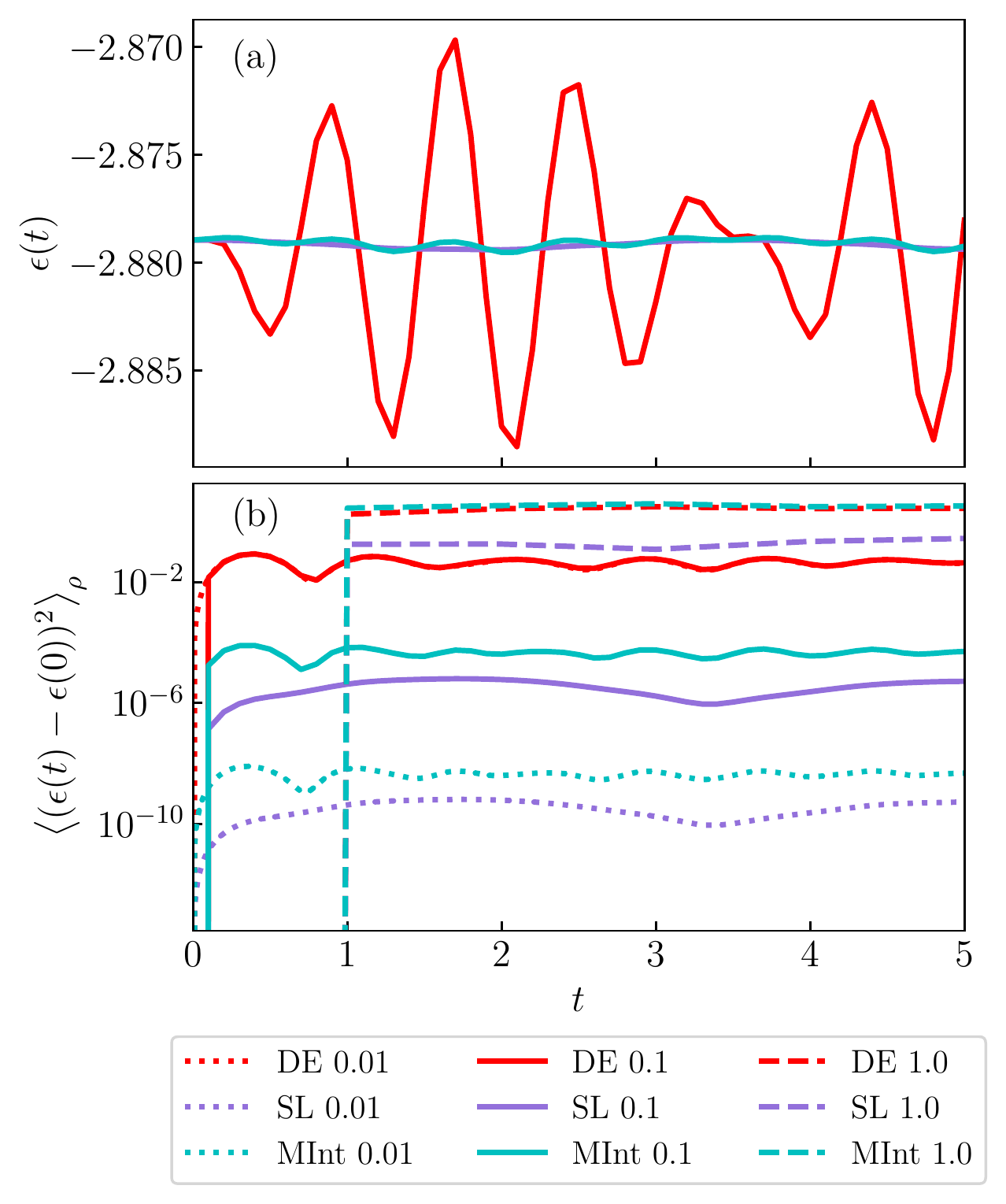}
\caption{The energy conservation for Model 1 with (a) a single trajectory and $\Delta t = 0.1$ (solid) and (b) averaged using $\Delta t = 0.1$ (solid), $\Delta t = 0.01$ (dotted) and $\Delta t = 1.0$ (dashed) with the SL (purple), the MInt (cyan), and the DE (red) algorithms. The DE algorithm has the worst energy conservation and, for this system, the MInt has slightly worse conservation than the SL. The SL and MInt algorithms show improved energy conservation upon decreasing the time step, unlike the DE.}
\label{fig2}
\vspace{-10pt}
\end{figure}

We now look at energy conservation, Fig.~\ref{fig2}, where (a) depicts the energy of a single trajectory and (b) averages the energy conservation criterion,
\begin{equation} \label{energyequation}
    \langle(\epsilon(t) - \epsilon(0))^2\rangle_{\rho} = \frac{\sum_{i=1}^J (\epsilon(t) - \epsilon(0))^2 W_{i}(0)}{\sum_{i=1}^J 
 W_{i}(0)} \text{,}
\end{equation} 
over trajectories until convergence was observed. We calculate the energy, $\epsilon$, by evaluating the MMST Hamiltonian, Eqn.~\ref{MMST}, at each timestep. Under perfect energy conservation, the criterion should be zero. For a single trajectory, we observe that the DE algorithm has much larger oscillations and does not conserve energy well when compared to the MInt and SL algorithms.
In Fig.~\ref{fig2}b, we average over many trajectories and consider different timestep sizes. It is seen that the MInt and SL algorithms are second order, as changing the timestep by a factor of 10 increases the criterion by $\sim 10^4$ as expected.\cite{Leimkuhler2005} This agrees with the algebraic determination of the order by Church \textit{et al.}.\cite{Church2018} However, the DE algorithm stays the same magnitude for $\Delta t = 0.01$ and $\Delta t = 0.1$, appearing to not be affected by the smaller timestep.
The very coarse $\Delta t = 1.0$ is so large that it breaks the trends.
The DE algorithm's poor energy conservation is due to the discarded terms when the DE approximation is made; when these terms are large, the propagation of the nuclear momentum is affected significantly which then affects all other variables through the propagation equations. 
The MInt and SL energy conservation is very similar although the SL algorithm is seen to have the smallest energy fluctuations throughout and appears to have a longer period of oscillation for Model 1.
This is surprising as one would expect a symplectic algorithm to have better energy conservation compared to a non-symplectic algorithm.  
For Models 2 and 3, we observe that the MInt and SL algorithms have almost identical energy conservation whilst the DE algorithm is worse, seen in Figs.~1 and 2 in the SI.

\subsection{Correlation Functions}
\begin{figure}[H]
\centering
\includegraphics[width=\textwidth]{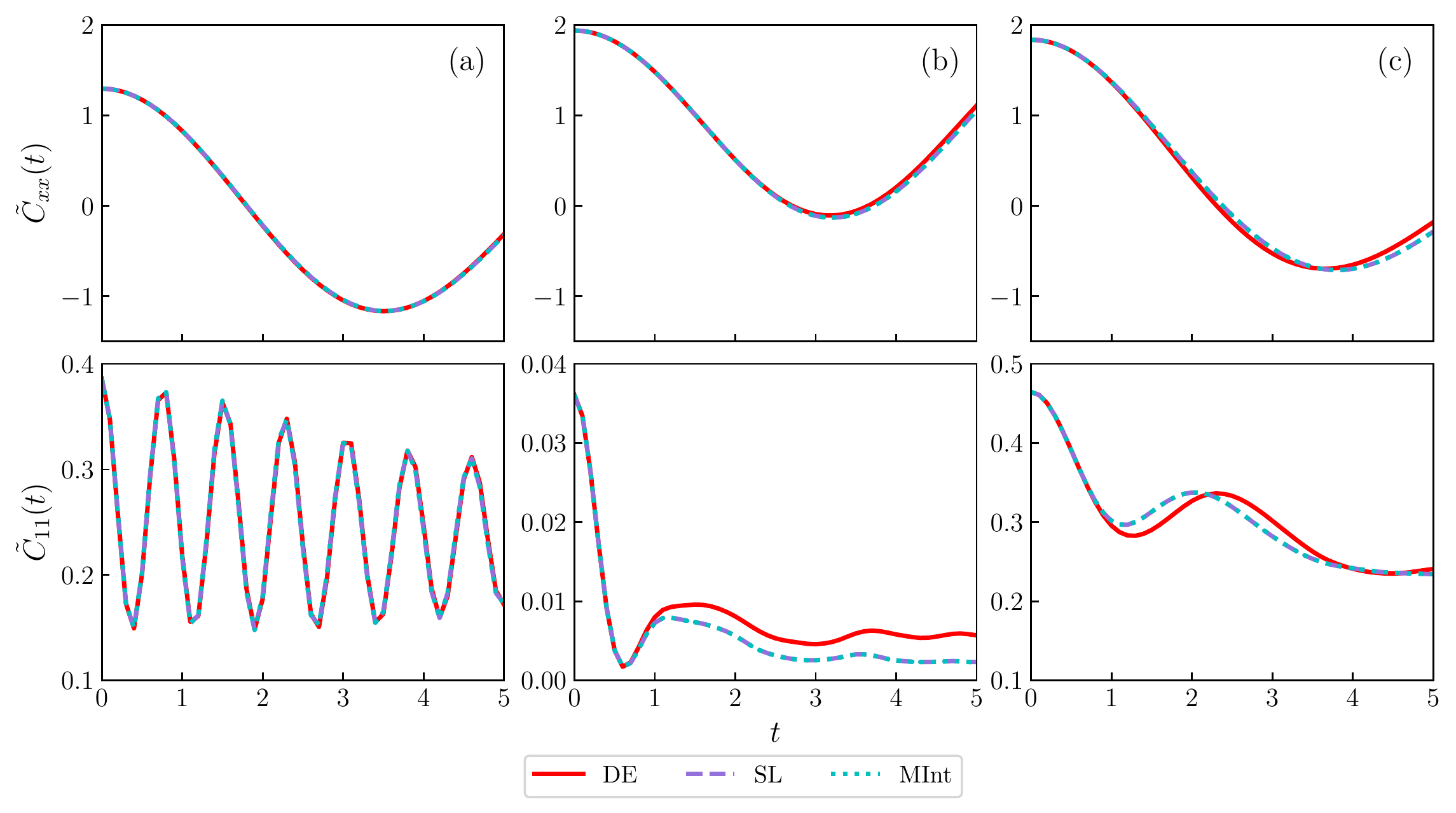}
\vspace{-10pt}
\caption{The nuclear position, $\tilde{C}_{xx}(t)$, and electronic population, $\tilde{C}_{11}(t)$, autocorrelation functions for (a) Model 1, (b) Model 2 and (c) Model 3
using $\Delta t = 0.1$ with the DE (red solid), SL (purple dashed) and MInt (cyan dotted) algorithms. The SL and MInt give identical results, but the DE algorithm deviates from them in (b) and (c).}
\label{fig3}
\end{figure}

The nuclear position and electronic population autocorrelation functions, Eqn.~\eqref{cfsamp}, were calculated for the three models. The fast oscillations of $\tilde{C}_{11}$ in Fig.~\ref{fig3}a indicate that the strong electronic coupling is close to the adiabatic limit and the dynamics tend towards standard RPMD as $U$ is approximately harmonic.\cite{Richardson2013a} For Models 2 and 3, the weaker electronic coupling reduces the electronic oscillation frequency, producing curves that deviate from the adiabatic result.\cite{Richardson2013a} In Model 2, the equilibrium population of the first electronic state is quickly lost to the lower energy second state, indicating that the system is almost always on one diabatic surface. The correlation functions obtained using the MInt and SL algorithms qualitatively replicate the dynamics expected from the single-bead calculation in Fig.~1 of Ref.~[\!\!\citenum{Richardson2013a}] for all models.  Comparing the three algorithms tested in Fig.~\ref{fig3} provides the interesting discovery that the DE algorithm is very accurate for Model 1, despite the lack of energy conservation. 
This is likely due to averaging with fast electronic oscillations providing the correct convergence. However, for the other models the DE algorithm predicts the same initial drop in the electronic population autocorrelation functions as the MInt and SL algorithms, but starts to deviate from the expected behaviour after the minima. The correlation functions have the same shape indicating a systematic error that likely arises due to the DE approximation. This assumes the off-diagonal elements of $\tilde{\bo{V}}'$ in the adiabatic basis, ${\bf{G}}$, are zero and that the diagonal elements are equal, which is not the case for the SL and MInt algorithms.\cite{Church2018}

Whilst testing energy conservation, we observed that the DE algorithm has more trajectories with poor energy convergence for models with weaker electronic coupling and was particularly poor in the intermediate regime given by Model 3. The MInt and SL algorithms produce the same correlation functions for all the models tested with small timesteps. When testing different timesteps, as seen in Appendix \ref{symp}, both the MInt and SL algorithms are tolerant of a coarse timestep. For $\Delta t = 1.0$, we observe that the MInt and SL correlation functions start to differ, with the MInt being closer to the small timestep results. 
However, the electronic oscillations are not captured well due to aliasing. The MInt and SL algorithms are limited by the model used rather than the algorithmic accuracy. 
\begin{center}
    \begin{tabular}{l|c|c|c}
      \toprule % <-- Toprule here
      \textbf{Computational Results} & \textbf{ MInt } & \textbf{ SL } & \textbf{ DE }\\
      \midrule % <-- Midrule here
      \rowcolor[HTML]{C0C0C0} Satisfies Liouville's Theorem & \cmark & \cmark & \cmark\\
      Symplectic & \cmark & \xmark & \xmark \\
      \rowcolor[HTML]{C0C0C0} Energy Conservation & Good & Good & Poor \\
      Correlation Function Accuracy & Good & Good & Poor \\
      \bottomrule % <-- Bottomrule here
    \end{tabular}
    \captionof{table}[Summary of the computational results that are exact and consistent with the theoretical results in Table \ref{theoreticaltable}.]{Summary of the computational results, which are consistent with the theoretical results in Table \ref{theoreticaltable}.}
    \label{table2}
\end{center}

In Table \ref{table2}, we provide an overview of the properties tested here. The MInt algorithm is the only symplectic algorithm satisfying the symplecticity criterion, providing exact propagation of $H_{1}$ and $H_{2}$.\cite{Church2018} The SL and DE algorithms have a non-zero error matrix that increases with time, indicating that neither are symplectic due to the approximations made.\cite{Church2018, Richardson2013a, Richardson2017, Kelly2012} The lack of symplecticity may
\end{multicols}
\begin{multicols}{2}
\noindent result in an upwards energy drift at long simulation times, although this was not observed within the short simulation time tested here. 

The SL algorithm approximates that the electronic variables can be held still while the nuclear momentum is propagated.\cite{Richardson2017} The DE algorithm assumes that the eigenvalues of the potential matrix are equal, such that the potential derivative in the adiabatic basis is approximated as $\bo{S}^\textrm{T}\tilde{\bo{V}}'\bo{S} = \boldsymbol{\Lambda}'$.\cite{Kelly2012} This results in inaccurate nuclear trajectories that leads to poor energy conservation. However, all three algorithms obey Liouville's Theorem, preserving volume phase-space throughout the trajectories. The MInt and SL have similar energy conservation and are both second-order algorithms, where in some cases the SL has better energy conservation.\cite{Church2018, Leimkuhler2005} 

The correlation functions produced for the MInt and SL algorithms are very similar to those computed in Ref.~[\!\!\citenum{Richardson2013a}], with the small differences arising from the different choice of splitting the potential. The DE algorithm converges to a different result for models with weaker coupling, being a good approximation only near the adiabatic limit.

Table 1 in the SI presents timings of the computational algorithms. One should note that in this case, the bottleneck is the evaluation of the algorithm itself rather than the calculation of the potential matrix.
When calculating the monodromy matrix, we observed that the SL algorithm has the lowest computational cost followed closely by the MInt, with the DE algorithm taking the longest time to run. Without calculating the monodromy matrix, the SL is significantly faster and the MInt and DE take similar times to run.

\section{Conclusions} \label{conclusions}
In this article we have tested symplecticity, Liouville's theorem, energy conservation and computed correlation functions using the MInt, SL and DE algorithms for a range of model parameters and timesteps. We find that the computational results agree with our theoretical predictions. If symplecticity is required, for accurate MMST Hamiltonian dynamics with little energy drift, the MInt algorithm should be used. As far as we are aware, the MInt is the only known symplectic algorithm for a general form of the MMST Hamiltonian. 
However, even though the SL algorithm is not formally symplectic with a finite timestep, it becomes exact in the limit of an infinitesimal timestep.
In our tests, it gave comparable accuracy to the MInt algorithm, but at a lower computational cost.
We would not recommend the DE algorithm for these models, as it breaks energy conservation and introduces errors into the results.  
This indicates that for the models used here, the degenerate eigenvalue approximation is not valid. 

Further work includes integrating the Cayley transform to extend these findings to stable NRPMD simulations.\cite{Korol2019} Additionally, one can apply the MInt algorithm to related dynamical methods such as forward-backward (FB)-IVR.\cite{Thoss2001, Malpathak2022}

\section*{Supporting Information} \label{SI}
The supplementary material comprises of: additional energy conservation plots for Models 2 and 3, comparison of correlation functions using different timesteps and the properties of the DE algorithm in the original form.  

\section*{Acknowledgements}
TJHH acknowledges a Royal Society University Research Fellowship URF\textbackslash R1\textbackslash 201502. LEC acknowledges a University College London studentship. JER acknowledges funding from the ETH Hans H. Günthard fellowship. 

%\begin{appendices}
\appendix
    \renewcommand{\thesubsection}{\Alph{subsection}}
    \section*{Appendices}
    \addcontentsline{toc}{section}{Appendices}
    \subsection{Symplecticity}
    \label{symp}
    \setcounter{equation}{0}
    \setcounter{figure}{0}
    \setcounter{table}{0}
    \numberwithin{equation}{subsection}
    \numberwithin{figure}{subsection}
    \numberwithin{table}{subsection}
    The symplecticity criterion can be algebraically determined for the three algorithms, where the MInt and SL have already been derived by Church \textit{et al.}~(2018). It is sufficient to say that under the splitting of the Hamiltonian into $H_{1}$ and $H_{2}$, the symplecticity criterion becomes\cite{Church2018, Leimkuhler2005} 
    \begin{equation}
        \bo{M}_{H_{1}}^\textrm{T}\bo{J}^{-1}\bo{M}_{H_{1}} = \bo{J}^{-1} \quad \text{and} \quad  \bo{M}_{H_{2}}^\textrm{T}\bo{J}^{-1}\bo{M}_{H_{2}} = \bo{J}^{-1}.
    \end{equation}
    For example, following the method by Church \textit{et al.}, $M_{H_{1}}$ is Eqn.~\eqref{H1Mono} for all three algorithms. This can be derived from the propagation equations, Eqn.~\eqref{H1HE}, where $\textrm{M}_{XX} = \textrm{M}_{xx} = \textrm{M}_{pp} = \textrm{M}_{PP}=1$ and
    \begin{equation}
        \label{Mxp}
        \textrm{M}_{xp} = \frac{\partial x} {\partial p} = \frac{\partial} {\partial p} \left( x(t) + \frac{\Delta t}{m}p \right) = \frac{\Delta t}{m}. 
    \end{equation}
    Evaluating the symplecticity criterion results in
    \begin{equation}
        \label{H1symp}
        \bo{M}_{H_1}^\textrm{T}\bo{J}^{-1} \bo{M}_{H_{1}}= \left[ {\begin{array}{cccc}
        0 & \bo{0}^\textrm{T} & -1 & \bo{0}^\textrm{T}  \\
        \bo{0}  & \mathbb{O} & \pmb{0} & -\mathbb{I} \\
        1 & \bo{0}^\textrm{T} & 0 & \bo{0}^\textrm{T}\\
        \bo{0}  &\mathbb{I} & \pmb{0} & \mathbb{O}  \\
      \end{array} } \right] = \bo{J}^{-1} \text{,}
    \end{equation}
    such that the propagation under $H_{1}$ is symplectic.\cite{Church2018} 
    
    The symplecticity for $H_{2}$ is derived separately for all three algorithms due to the different nuclear propagation equations, where the MInt and SL algorithms had previously been derived by Church \textit{et al.}. For the MInt algorithm, $M_{H_{2}}$ is Eqn.~\eqref{H2monomint} where the symplecticity criterion becomes
    
    \begin{equation}
        \label{MIntH2symp}
        \bo{M}_{H_2}^\textrm{T}\bo{J}^{-1} \bo{M}_{H_{2}}= \left[ {\begin{array}{cccc}
        0 & \bo{h}^\textrm{T}  & -1 & \bo{j}^\textrm{T}  \\
        \bo{h}  & \mathbb{O} & \pmb{0} & -\mathbb{I}\\
        1 & \bo{0}^T & 0 & \bo{0}^T\\
        \bo{j}  &\mathbb{I} & \pmb{0} & \mathbb{O} \\
      \end{array} } \right] \text{,}
    \end{equation}

   \noindent \hrulefill
    
    with
    \begin{subequations} \label{hj} \begin{align}
        \bo{h} &= -\bo{Cf} +\bo{e}^\textrm{T}+\bo{Dg} \text{,} \\
        \bo{j} &= \bo{Df} +\bo{a}^\textrm{T} +\bo{Cg} \text{,}
        \end{align}
    \end{subequations}
    where Church \textit{et al.}~previously determined that $\bo{h} \equiv \bo{0}$ and $\bo{j} \equiv \bo{0} \ \forall \ \bo{X}, \bo{P}$.
    
    For the SL algorithm, propagation of $H_{2}$ is split further into a nuclear and electronic contribution for which the monodromy matrices are Eqn.~\eqref{H2monosl}. Evaluating the symplecticity criterion for $\bo{M}_{\bo{p}}$
    \begin{equation}
        \label{Mpsymp}
        \bo{M}_{\bo{p}}^\textrm{T}\bo{J}^{-1} \bo{M}_{\bo{p}}= \left[ {\begin{array}{cccc}
        0 & \frac{\Delta t}{2}\bo{X}^\textrm{T}\tilde{\bo{V}}'  & -1 & \frac{\Delta t}{2}\bo{P}^\textrm{T}\tilde{\bo{V}}'  \\
        -\frac{\Delta t}{2}\tilde{\bo{V}}'\bo{X} & \mathbb{O} & \pmb{0} & -\mathbb{I}\\
        1 & \bo{0}^\textrm{T} & 0 & \bo{0}^\textrm{T}\\
        -\frac{\Delta t}{2}\tilde{\bo{V}}'\bo{P}  &\mathbb{I} & \pmb{0} & \mathbb{O} \\
      \end{array} } \right] \text{,}
    \end{equation}
    which will only be symplectic if $\tilde{\bo{V}}' = 0$.\cite{Church2018} Evaluating for $\bo{M}_{\textrm{el}}$
    \begin{align}
        \label{Helsymp}
        &\bo{M}_{\textrm{el}}^\textrm{T}\bo{J}^{-1} \bo{M}_{\textrm{el}} \\ & = \left[ {\begin{array}{cccc}
        0 & -\bo{g}^\textrm{T}\bo{D}+\bo{f}^\textrm{T}\bo{C}  & -1 & -\bo{g}^\textrm{T}\bo{C}-\bo{f}^\textrm{T}\bo{D}   \\
        -\bo{Cf} +\bo{Dg} & \mathbb{O} & \pmb{0} & -\mathbb{I}\\
        1 & \bo{0}^\textrm{T} & 0 & \bo{0}^\textrm{T}\\
        \bo{Df} +\bo{Cg}  &\mathbb{I} & \pmb{0} & \mathbb{O} \\
      \end{array} } \right] \nonumber \\ &= \left[ {\begin{array}{cccc}
        0 & \bo{e}  & -1 & \bo{a}  \\
        -\bo{e}^\textrm{T} & \mathbb{O} & \pmb{0} & -\mathbb{I}\\
        1 & \bo{0}^T & 0 & \bo{0}^T\\
        -\bo{a}^\textrm{T} &\mathbb{I} & \pmb{0} & \mathbb{O} \\
      \end{array} } \right] \text{,} \nonumber
    \end{align}
    where Eqn.~\eqref{hj} has been used in conjunction with the fact that $\bo{h} \equiv \bo{0}$ and $\bo{j} \equiv \bo{0} \ \forall \ \bo{X}, \bo{P}$.\cite{Church2018} 
    
    Church \textit{et al.}~note that $\bo{a}, \bo{e} \ne \bo{0}^\textrm{T}$ so evolution under $\mathscr{L}_{\textrm{el}}$ and $\mathscr{L}_{\bo{p}}$ is not symplectic. The combined evolution under $\mathscr{L}_{\textrm{el}}$ and $\mathscr{L}_{\bo{p}}$ does not provide error cancellation that restores symplecticity. We find that numerically the difference between $\bo{M}_{H_{2}}$ and $\bo{M}_{{\textrm{el}}}\bo{M}_{{\bo{p}}}$ is on $\mathcal{O}(\Delta t)^2$, in agreement with the literature.\cite{Church2018} The combination of $\bo{M}_{{\bo{p}}}\bo{M}_{{\textrm{el}}}$ also provides the same result. This means the propagation of $\bo{M}_{{\bo{p}}}\bo{M}_{{\textrm{el}}}$ will be symplectic in the $\Delta t \to 0$ limit, but will not be for an arbitrary timestep. 
    
    For the DE algorithm, the monodoromy matrix, $\bo{M}_{\textrm{DE}}$,  for the approximate propagation of $H_{2}$ has been derived for the first time as Eqn.~\eqref{Mde}.  
    Using Eqn.~\eqref{monoelements} with $\bo{h} \equiv \bo{0}$ and $\bo{j} \equiv \bo{0} \ \forall \ \bo{X}, \bo{P}$, the symplecticity criterion becomes
    \end{multicols}
    \renewcommand{\thesubsection}{\Alph{subsection}}
    \begin{align}
        \label{DEsymp}
        \bo{M}_{\textrm{DE}}^\textrm{T}\bo{J}^{-1} \bo{M}_{\textrm{DE}} &= 
        \left[ {\begin{array}{cccc}
        0 & \bo{e} + \frac{\Delta t}{2}\bo{X}^\textrm{T}\bo{S}\boldsymbol{\Lambda}'\bo{S}^\textrm{T} & -1 & \bo{a} + \frac{\Delta t}{2}\bo{P}^\textrm{T}\bo{S}\pmb{\Lambda}'\bo{S}^\textrm{T} \\
        \bo{e}^\textrm{T} - \frac{\Delta t}{2}\bo{S}\pmb{\Lambda}'\bo{S}^\textrm{T}\bo{X} & \mathbb{O} & \pmb{0} & -\mathbb{I}\\
        1 & \bo{0}^\textrm{T} & 0 & \bo{0}^\textrm{T}\\
        \bo{a}^\textrm{T} - \frac{\Delta t}{2}\bo{S}\pmb{\Lambda}'\bo{S}^\textrm{T}\bo{P} &\mathbb{I} & \pmb{0} & \mathbb{O} \\ 
      \end{array} } \right] \text{,}
    \end{align}
    \begin{multicols*}{2}
    
    \renewcommand{\thesubsection}{\Alph{subsection}}
    \noindent which will not be symplectic for the same reasoning as for the SL algorithm. 
    
    To derive the order of the difference between $\bo{M}_{H_{2}}$ and $\bo{M}_{\textrm{DE}}$, following the approach of Church \textit{et al.}~, we define

\begin{figure}[H]
    \centering
    \includegraphics[width=0.49\textwidth]{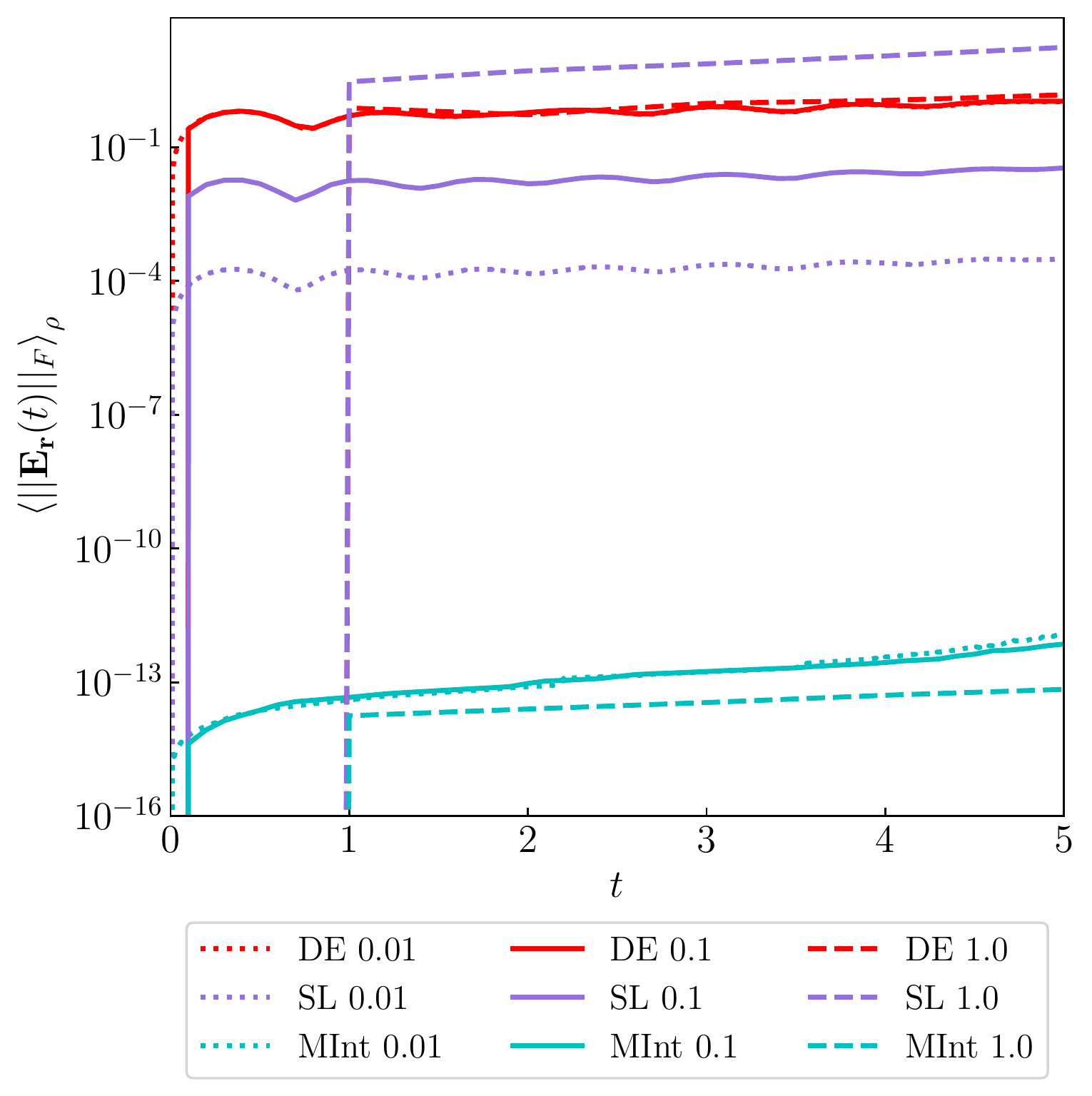}
    \caption{The Frobenius Norm of the error matrix averaged until convergence with the SL (purple), the MInt (cyan), and the DE (red) algorithms using timesteps of $\Delta t = 0.1$ (solid), $\Delta t = 0.01$ (dotted) and $\Delta t = 1.0$ (dashed). The MInt algorithm is seen to be symplectic and does not change with timestep. The SL algorithm is seen to be second order with respect to time and the DE algorithm is zero order, in agreement with the theoretical predictions.}
    \label{figA1}
    \end{figure}

    \noindent \hrulefill 
    \begin{subequations} 
        \label{deAE} \begin{align}
        \overline{\bo{a}} = - \frac{\Delta t}{2}\bo{S}\pmb{\Lambda}'\bo{S}^\textrm{T}\bo{X} \text{,} \\
        \overline{\bo{e}} = - \frac{\Delta t}{2}\bo{S}\pmb{\Lambda}'\bo{S}^\textrm{T}\bo{P} \text{,} \end{align}
    \end{subequations}
    such that, the symplecticity criterion will be met if $\overline{\bo{a}} \equiv \bo{a}$ and $\overline{\bo{e}} \equiv \bo{e}$. Expanding in coefficients of {\bf{X}} and {\bf{P}} leads to
    \begin{subequations} 
        \label{EFequiv} \begin{align}
        {\bf{E}} & \stackrel{?}{=}  \frac{ \Delta t}{2}\bo{S}\pmb{\Lambda}'\bo{S}^\textrm{T} \text{,} \\
        {\bf{F}} & \stackrel{?}{=} \mathbb{O} \text{,} \end{align}
    \end{subequations}
    Rotating to the diabatic basis gives 
    \begin{subequations} 
        \label{EFequivdiabatic} \begin{align}
        {\bf{S}}^T \left( {\bf{E}} - \frac{ \Delta t}{2}\bo{S}\pmb{\Lambda}'\bo{S}^\textrm{T} \right){\bf{S}} &= {\boldsymbol{\Gamma}} - \frac{ \Delta t}{2}\pmb{\Lambda}' \text{,} \\
        {\bf{S}}^\textrm{T} {\bf{F}}{\bf{S}}  & = {\boldsymbol{\Xi}} \text{,} \end{align}
    \end{subequations}
    where ${\boldsymbol{\Gamma}}$ and ${\boldsymbol{\Xi}}$ are given in Eqn.~\eqref{gammaxi}.
    Evaluating element-wise in powers of $\Delta t$ 
    \begin{subequations} 
        \label{deltatequiv} \begin{align}
        ({{\boldsymbol{\Gamma}} - \frac{ \Delta t}{2}\pmb{\Lambda}'})_{nm} &= \begin{cases} 
        0 &\ n=m\\
        -({\bf{S}}^\textrm{T}{\bf{S}}')_{nm} \left(\lambda_{nm}\frac{\Delta t}{2} + \mathcal{O}(\Delta t)^3 \right) &\ n \neq m \text{,}
        \end{cases} \\
        {\boldsymbol{\Xi}}_{nm} &= \begin{cases} 
        0  &\ n=m\\
        \frac{\lambda_{nm}}{2} (\frac{\Delta t}{2} )^2 {\bf{G}}_{nm} + \mathcal{O}(\Delta t)^4 &\ n \neq m \text{.}
        \end{cases}  \end{align}
    \end{subequations}
    Hence, the order of the difference between $\bo{M}_{H_{2}}$ and $\bo{M}_{\textrm{DE}}$ is $\mathcal{O}(\Delta t)$. The SL algorithm, $\bo{M}_{\textrm{el}}\bo{M}_{\bo{p}}$, is one order higher than the DE algorithm, $\bo{M}_{\textrm{DE}}$, so the symplecticity is better as seen in Figs.~\ref{fig1} and \ref{figA1}.

    We find that, numerically, when considering the whole propagation of $H$ using the SL algorithm, the initial difference in symplecticity between the MInt and SL algorithms scale on order of $\mathcal{O}(\Delta t)^3$. This is due to some error cancellation with the symmetric splitting of $H_{2}$.  Hence, when propagating for a given length of time, the order is $\mathcal{O}(\Delta t)^3/ \Delta t = \mathcal{O}(\Delta t)^2$, seen in Fig.~\ref{figA1}. For the DE algorithm, the initial difference in symplecticity with the MInt algorithm scales on $\mathcal{O}(\Delta t)$. Therefore, for a given length of time, the order is $\mathcal{O}(\Delta t)/ \Delta t = 1$. Fig.~\ref{figA1} corroborates this as the DE algorithm appears to be on zero order for all the timesteps tested. The symplecticity does not improve with the smaller timesteps. 

    \renewcommand{\thesubsection}{\Alph{subsection}}
    \subsection{Sampling}
    \label{samp}
    Following a similar derivation of the partition function as in Refs.~[\!\!\citenum{Richardson2017}] and [\!\!\citenum{Richardson2013a}], the sampling method used in this work is derived. Instead of formulating in the limit of an infinite number of ring-polymer beads, we derive the partition function in the single-bead case. The wavefunctions of the singly-excited oscillator (SEO) states are known in the position and momentum bases\cite{Richardson2017} 
    \begin{align}
        \label{SEOwavefunctions}
        \braket{\bf{X}}{n} = \sqrt{\frac{2}{\pi}}X_{n} e^{-|{{\bf{X}}^2}|/2} \text{,} \\ \braket{\bf{P}}{n} = -i\sqrt{\frac{2}{\pi}}P_{n} e^{-|{{\bf{P}}^2}|/2} \text{,}
    \end{align}
    where $\braket{n}{\bf{X}} = \braket{\bf{X}}{n}^*$ and likewise for \e{P}. The partition function, $Z=\text{Tr}[e^{-\beta \hat{H}}]$, is expanded as a Trotter product where $N=1$ for the one bead case
    \begin{align}
        \label{partitionbead}
        Z \simeq \text{Tr} \left[\prod_{i=1}^N e^{-\beta \frac{p^2}{2m}}e^{-\beta{\bf{V}}(x)} \right] \text{.}
    \end{align}
    Inserting the projection operator $\mathcal{P} = \sum_{n=1}^2 \ket{n}\bra{n}=1$ and summing over the two electronic levels in our model
    \begin{align}
        \label{partition1} \begin{split}
        Z \simeq \text{Tr} \Biggl[\sum_{k,l,m,n=1}^2 \int_{-\infty}^{\infty} \ket{k}\braket{k}{\bf{X}}\braket{\bf{X}}{l}\bra{l}e^{-\beta {\bf{V}}/2}\ket{x} \\ \times \bra{x} e^{-\beta \frac{p^2}{2m}}\ket{m}\braket{m}{\bf{P}}\braket{\bf{P}}{n} \\ \times \bra{n}e^{-\beta {\bf{V}}/2} \ \mathrm{d}x\mathrm{d}{\bf{X}}\mathrm{d}{\bf{P}}\Biggr] \text{,} \end{split} 
    \end{align}
    where the imaginary-time free-particle propagator for $N=1$ is\cite{Richardson2017} 
    \begin{equation} \label{imtimefreeparticle}
    \bra{x} e^{-\beta \frac{p^2}{2m}}\ket{x} =\sqrt{\frac{m}{2\pi\beta\hbar^2}} \text{,}
    \end{equation}
    such that
    \begin{align}
        \label{partition2}
        \begin{split} Z \simeq \text{Tr} \Biggl[\sum_{k,l,m,n=1}^2 \int_{-\infty}^{\infty} \frac{4}{\pi^2}\sqrt{\frac{m}{2\pi\beta\hbar^2}} e^{-|{{\bf{X}}^2}| -|{{\bf{P}}^2|}} \\ \times  X_{k}X_{l}\ket{k} \bra{l}e^{-\beta {\bf{V}}/2}\ket{m} \\ \times  \bra{n} P_{m}P_{n}e^{-\beta {\bf{V}}/2} \ \mathrm{d}x\mathrm{d}{\bf{X}}\mathrm{d}{\bf{P}}\Biggr]. \end{split}
    \end{align}
    Separating out the state-independent potential, $U$,
    \begin{align}
        \label{partition3}
        \begin{split} Z \simeq \text{Tr} \Biggl[\sum_{k,l,m,n=1}^2 \int_{-\infty}^{\infty} \frac{4}{\pi^2}\sqrt{\frac{m}{2\pi\beta\hbar^2}} e^{-|{{\bf{X}}^2}| -|{{\bf{P}}^2|} -\beta{U}} \\ \times  X_{k}X_{l}\ket{k}\bra{l}e^{-\beta \tilde{{\bf{V}}}/2}\ket{m}\bra{n} \\ \times  P_{m}P_{n}e^{-\beta\tilde{{\bf{V}}}/2} \ \mathrm{d}x\mathrm{d}{\bf{X}}\mathrm{d}{\bf{P}}\Biggr] \text{.} \end{split}
    \end{align}
    The sum can be rewritten in matrix notation  
    \begin{align}
        \label{partitionmatrix}
        \begin{split} Z \simeq \int_{-\infty}^{\infty} \frac{4}{\pi^2}\sqrt{\frac{m}{2\pi\beta\hbar^2}} e^{-|{{\bf{X}}^2}| -|{{\bf{P}}^2|} -\beta{U}} \times W \ \mathrm{d}x\mathrm{d}{\bf{X}}\mathrm{d}{\bf{P}} \text{,} \end{split}
    \end{align}
    where when ${\bf{M}}=e^{-\beta{\tilde{\bf{V}}}/2}$ and using the cyclic properties of the trace 
    \begin{align}
        \label{Wcyclic}
        \begin{split} W = \text{Tr} \Biggl[ {\bf{P}}^\textrm{T}{\bf{MX}\bf{X}}^\textrm{T} {\bf{MP}} \Biggr] = {\bf{P}}^\textrm{T}{\bf{X}\bf{MX}}^\textrm{T} {\bf{MP}} \text{,} \end{split} 
    \end{align}
    which is a scalar quantity. Using the Gaussian integral identity 
    \begin{align}
        \label{gaussianintegral}
        \begin{split} \int_{-\infty}^{\infty} e^{-\beta p^2/2m} dp = \sqrt{\frac{2m\pi}{\beta}} \text{,} \end{split}
    \end{align}
    such that $p$ can be inserted into the integral 
    \begin{align}
        \label{partitionallvariables}
        \begin{split} Z \simeq \frac{1}{2\pi\hbar}\int_{-\infty}^{\infty} \frac{4}{\pi^2} e^{-|{{\bf{X}}^2}| -|{{\bf{P}}^2|} -\beta(U + p^2/2m)} \\ \times {\bf{P}}^\textrm{T}{\bf{MX}\bf{X}}^\textrm{T} {\bf{MP}} \   \mathrm{d}x\mathrm{d}p\mathrm{d}{\bf{X}}\mathrm{d}{\bf{P}} \text{,} \end{split}
    \end{align}
    resulting in the probability distribution, $\rho$, Eqn.~\eqref{distribution}, where $W$ is positive-definite in the single-bead case 
    \begin{align}
        \label{positivedefiniteW}
        \bo{M}^\textrm{T} = e^{-\beta{\tilde{\bf{V}}}^\textrm{T}/2}  = e^{-\beta{\tilde{\bf{V}}}/2}  = \bo{M}  \text{,}
    \end{align}
    therefore 
    \begin{align}
        \label{Wsquared}
        ({\bf{P}}^\textrm{T}{\bf{MX}})^\textrm{T} = {\bf{X}}^\textrm{T}{\bf{M}}^\textrm{T}({\bf{P}}^\textrm{T})^\textrm{T} = {\bf{X}}^\textrm{T}{\bf{MP}}.
    \end{align}
    This means that $W = ({\bf{X}}^\textrm{T}{\bf{MP}})^2$, which will always be positive. Sampling from Gaussian distributions for electronic variables where $\mu = 0$ and $\sigma =1/ \sqrt{2}$, and classical distributions for nuclear variables where $\mu_x = 0$ and  $\sigma_x = \sqrt{1 / (m\omega^2\beta)}$, and $\mu_p = 0$ and $\sigma_p = \sqrt{m /\beta}$, such that the sampled distribution is  
    \begin{align}
        \label{sampleddistribution}
        \rho_{\textrm{samp}} =  \frac{\beta \omega}{2\pi ^2} e^{-|{\bf{X}}^2|-|{\bf{P}}^2|-\beta(U + p^2 / 2m)} \text{.}
    \end{align}
    To obtain the required distribution, Eqn.~\eqref{distribution}, the following relationship is established
    \begin{align}
        \label{samptorequired}
        \rho = \rho_{\textrm{samp}} \times \frac{8
        %|W|
        }{\beta\omega} \text{,}
    \end{align}
    Hence, we need to correct observables calculated to average over the required distribution. %Letting $\rho = ( 4/ \pi^2) e^{-Z}$,  where $Z = |{\bf{X}}^2|+|{\bf{P}}^2|+\beta(U + p^2/ 2m) $,
    We can thus define the average of an observable over the distribution, $\hat{A}$, that may be a function of initial position and momenta i.e, $\hat{A}({\bf{X}}_{i},{\bf{P}}_{i}, x_{i},p_{i})$ as 
    \begin{align}
        \label{observableint} \begin{split}
        \langle A \times W \rangle_{\rho} = \int_{-\infty}^{\infty} A \times W \times \frac{\rho}{\rho_{\textrm{samp}}} \times \rho_{\textrm{samp}}  \ \mathrm{d}x\mathrm{d}p\mathrm{d}{\bf{X}}\mathrm{d}{\bf{P}}.
        \end{split} 
    \end{align}
    The integral over the sampled distribution can be replaced with a sum over the number of samples taken, $J$, 
    \begin{align}
        \label{observable}
        \langle A \times W \rangle_{\rho} &\simeq \frac{1}{J}\sum_{i=1}^J A \times W \times %\frac{\frac{4}{\pi^2}e^{-Z}}{\frac{\beta\omega}{2\pi^2}e^{-Z}} 
        \frac{\rho}{\rho_{\textrm{samp}}}\\ \label{observablefinal} & \simeq \frac{1}{J}\sum_{i=1}^J A \times W \times \frac{8}{\beta\omega}.
    \end{align}
    The partition function can be calculated as 
    \begin{align}
        \label{partitionsampledfinal}
         Z \propto \langle W \rangle_{\rho} &\propto \int_{-\infty}^{\infty} W \times {\rho} \  \mathrm{d}x\mathrm{d}p\mathrm{d}{\bf{X}}\mathrm{d}{\bf{P}} \nonumber \\ &\propto \frac{1}{J} \sum_{i=1}^J W \times \frac{8}{\beta\omega} \text{.}
    \end{align}
\addcontentsline{toc}{section}{References}
\bibliography{rp}
\bibliographystyle{rsc}

\end{multicols*}

\end{document}